\documentclass[12pt]{article}
\usepackage{amsfonts}
\usepackage{amsmath}
\usepackage{geometry}
\usepackage{longtable}
\usepackage{graphicx}
\usepackage{pdflscape}
\usepackage{setspace}

\newtheorem{theorem}{Theorem}

\newtheorem{assumption}{Assumption}

\newtheorem{lemma}{Lemma}

\newtheorem{proposition}{Proposition}
\newtheorem{remark}{Remark}

\newenvironment{proof}[1][Proof]{\noindent\textbf{#1.} }{\ \rule{0.5em}{0.5em}}
\begin{document}

\title{Demand and Welfare Analysis with Hedonic Budget Constraints}
\author{Debopam Bhattacharya\thanks{%
Financial support from Economic and Social Research Council (ESRC) is
acknowledged. We are grateful to Weida Liao for helpful conversations
regarding partial differential equations, and to Whitney Newey for feedback.
Email for correspondence: debobhatta@gmail.com} \\
%EndAName
University of Cambridge \and Ekaterina Oparina \\
%EndAName
LSE \and Qianya Xu \\
%EndAName
University of Cambridge}
\date{August 31, 2026 \\
\vspace{-1cm}}
\maketitle

\begin{abstract}
We analyze demand in hedonic settings where heterogeneous consumers maximize
utility for product attributes subject to smooth, nonlinear budget frontier.
We develop econometric methods for welfare-analysis of a change in the
frontier. Two key findings are Roy's identity for indirect utility yielding
a Partial Differential Equation system and Slutsky symmetry for quantile
demand, both holding under general heterogeneity. Further, under scalar
heterogeneity and single-crossing preferences, the coefficient functions in
the PDEs are shown to be point-identified. Under symmetry, these lead to
path-independent, money-metric welfare. We showcase our methods with a
simple empirical illustration using British microdata on home-rents and
neighborhood school quality.


\end{abstract}

\section{Introduction}

Hedonic budget constraints typically arise in modelling demand for
differentiated goods with a large number of available varieties, but where
each variety can be viewed as a distinct bundle of a limited number of
attributes or amenities (Rosen, 1974). Examples include buying cars, houses,
renting hotels, hiring on the labour market, etc. An important
characteristic of such markets is that in equilibrium, the marginal price of
an amenity typically varies with quantity, making budget frontiers nonlinear
(Diewert, 2003; Ekeland et al 2004). For example, Pakes 2003 finds the price
of personal computers to be increasing and concave in CPU speed. In
empirical models of labour supply, the income tax rate is often progressive,
causing potential workers to face piecewise linear budget constraints.
Indeed, if workers correctly viewed the tax schedule, one would expect to
observe bunching at kink points, which is often not observed in the data
(cf. Saez 2010). Hence MaCurdy et al. (1990, Section II.D) propose replacing
the piecewise linear budget-constraints with a smooth budget frontier to
reflect optimization and/or measurement error. The wage paid to workers with
specific skill-sets is conveniently modeled through hedonics since the
number of potential workers vastly outnumbers the number of key essential
skills (cf. Rosen and Sanderson 2002). Similarly, families choosing which
neighbourhood to live in consider the quality of local schools, whence the
potential choice-set becomes much larger than the key visible attributes of
schools viz. test-scores, student-teacher ratio and classification by
regulatory bodies such as the Ofsted in the UK and state education
departments in the United States. In such settings, explicit policy
interventions or market shocks can alter the equilibrium relationship
between product price and product attributes. For example, a relaxation of
catchment area restrictions in school choice is likely to weaken the
dependence of the equilibrium house-price on neighbourhood school-quality.
For individual households, this would lead to a flattening of their
perceived budget constraint and potentially change their choice and
resulting utility. The present paper develops econometric methods for
welfare analysis in such settings without specifying any arbitrary
functional form for \textit{unobserved} preference heterogeneity but
imposing a flexible parametric structure on the \textit{observed} hedonic
price schedule.

We begin our analysis with an economic model of choice for a heterogeneous
population of consumers facing a convex budget set characterized by a smooth
but nonlinear frontier in equilibrium. We consider an eventual change in
this frontier brought about by a policy change or a market shock, and seek
to quantify the resulting welfare effect on consumers. To do this, we first
derive an analog of Roy's Identity for this setting, which yields \textit{a
system} of linear partial differential equations (PDEs), unlike the standard
Roy's identity for two goods, which yields a single PDE. This result enables
the fundamental task of inferring welfare-relevant unobserved features of
consumer preferences from observable demand data. Next, we show that the
structural demand functions resulting from nonlinear budgets must satisfy a
Slutsky like symmetry condition which is different from the standard
textbook case of linear budgets with homogeneous consumers. The Slutsky
symmetry is a consequence of fundamental economic restrictions on behavior
and guarantees that welfare effects of budget changes are well-defined in
the sense of being path-independent. The above results are of independent
interest in the context of demand analysis with unrestricted heterogeneity,
pioneered by McFadden-Richter 1990 and reintroduced in econometrics by
Kitamura and Stoye 2013 and Hausman and Newey 2016. Next, we show that when
in particular unobserved heterogeneity is a scalar, and a single-crossing
condition is satisfied by preferences, the coefficient functions of the PDEs
resulting from Roy's identity can be identified from quantiles of demand.
Furthermore, welfare at the quantile can be expressed as a line integral
whose value is path-independent under our new Slutsky symmetry condition for
nonlinear budgets. These steps can be repeated separately for each quantile
of unobserved heterogeneity to obtain the entire distribution of welfare
effects. We emphasize that our goal here is to provide a methodology for
welfare calculation, once the \textit{equilibrium} budget frontiers have
been identified. This means that the equilibrium relation between price and
the amenity of interest is assumed to be consistently estimable both before
and after the change we are evaluating. In particular, we do not explicitly
discuss well-known issues of omitted variable problems that jeopardize
identification of the price-attribute relationship, on which there is a
significant literature in economics cf. Black 1999, Machin and Salvanes 2016
etc. In our empirical illustration below, we use an instrumental variable
based strategy to estimate this relationship to address potential
endogeneity resulting from omitted variables (see Sec \ref{Endog}).

As a motivating example, consider the well-known relationship between
housing costs and neighborhood school quality (Sheppard, 1999). To enable
children to attend nearby schools, local governments often mandate
`catchment area' rules, which restrict school access solely to neighborhood
children. This, however, means that the presence of a good school makes its
adjoining residential neighborhood attractive, and raises housing costs.
This leads to wealthier families moving in from worse school districts,
aggravating existing socioeconomic segregation. One potential way to stop
this vicious cycle is to relax catchment area restrictions. This would lead
housing costs to become less entangled with school quality, and change
choices in equilibrium. For example, Machin and Salvanes 2016 find that
relaxing catchment area boundaries in Oslo caused significant weakening of
the price-school-quality hedonic relation. However, the overall welfare
effects of such policy interventions are likely to be heterogeneous,
depending on both household preferences over consumption and neighborhood
school quality, and on their income. The question is: how can we calculate
the distribution of these heterogeneous welfare effects using microdata on
housing and schools. The present paper provides a theoretical framework and
corresponding econometric methodology to achieve this objective. The same
methodology would be applicable for studying the effect on homeowners when
improvement in transport links to the central business district changes the
equilibrium relationship between house price and house location, or the
impact on workers when the equilibrium wage-education relationship changes
in the labour market due to AI adoption or changes in immigration rules, or
the impact of technological advances in markets for consumer electronics,
such as laptops or mobile phones. In all these examples, the number of key
attributes affecting consumer choice are typically much smaller than the
number of potential choices, making these cases ideal for modelling via
hedonics, and thus amenable to the methods developed in the present
paper.\smallskip

\textbf{Contributions and Relation with Previous Literature}: Our key
theoretical contributions are the derivation of (a) an analog of Roy's
identity, and (b) an analog of Slutsky symmetry for structural demand, which
hold for nonlinear budget frontiers under unrestricted consumer
heterogeneity. For the canonical case of linear budget lines and a single
consumer, these two fundamental results feature in every undergraduate
textbook on microeconomics; however, the corresponding forms under nonlinear
budgets were previously unknown, to the best of our knowledge. Further, our
paper belongs to the tradition of empirical demand analysis with consumers
whose preferences are characterized by unobservable preference
heterogeneity, first discussed in Lewbel 2001. More specifically, our work
is substantively related to Heckman et al. 2010, who show how to identify
consumers' marginal utility for a continuous amenity in a hedonic setting,
where unobserved heterogeneity is a scalar, preferences are quasilinear in
consumption and satisfy a single-crossing condition. For the present paper,
we use the set-up in Heckman et al. (2010), except that utilities are not
assumed to be quasilinear in consumption, and our focus is on welfare
effects, which we show to be obtainable \textit{without} identifying the
underlying marginal utilities, the focus of Heckman et al. 2010.
Quasilinearity makes welfare calculations trivial by equating Marshallian
and Hicksian demand, but has the implication that demand is unaffected by
income (cf. Mas-Colell et al 1995, Page 83), which is unrealistic in
settings like home rental and purchase. In contrast to Heckman et al 2010,
however, we assume a flexible parametric form for the \textit{observable}
hedonic budget frontier in the main body of our paper. We do this both
because (a) it makes the parallel with the textbook case of linear budgets
transparent, and (b) even if we establish theoretical results keeping the
budget frontier nonparametric, for empirical implementation using a sieve to
approximate the frontier, we will need the results corresponding to the
parametric form. Nonetheless, for completeness, we outline in Appendix 2 the
analogous forms of Roy's Identity and Slutsky symmetry when the budget
frontier is left fully nonparametric.

Our paper continues a line of research started by the seminal article of
Hausman 1981, subsequently refined in Hausman and Newey 2016, who show that
in a demand setting with \textit{linear} budget frontiers, general
heterogeneity and income-effects, the welfare distribution resulting from a
price change are generically not point-identified. In contrast, our setting
allows hedonic budget frontiers to be smooth and nonlinear but we maintain
the Heckman et al 2010 assumption that unobserved heterogeneity is scalar
and satisfies a single crossing condition. We later show how to include
additional attributes into the analysis. Blomquist and Newey 2002 have shown
how to nonparametrically identify the distribution of demand on hypothetical 
\textit{linear} budget sets when observed budget constraints are continuous,
with the slope changing at finitely many known kink points. Once demand
distribution is identified for hypothetical \textit{linear} budget
constraints, welfare analysis would resort exactly to methods developed in
Hausman and Newey 2016. The presence of kinks play an important role in the
identification analysis of Blomquist and Newey et al 2002, as it defines the
budget set in terms of countably finite distinct intercepts, kinks and
slopes. In contrast, the general hedonic budget set considered in our set-up
is smooth; thus there are uncountably infinite slopes and intercepts and no
kinks. Hence the identification methods of Blomquist and Newey 2002 that
rely fundamentally on the finitely many kinks (see their theorems 2.1 and
2.2) do not apply here. More recently, Gan et al 2015 investigated welfare
calculation in fully \textit{parametric} models of labour supply (contrary
to the title of their paper), where the functional form of the indirect
utility function is specified up to finite dimensional parameters and
distribution of additive unobserved heterogeneity is assumed to be Gaussian
(cf. Equation (10), (11), (14) in Gan et al 2015). Identification of
preference parameters comes from a Tobit type likelihood function (eqn. (16)
in Gan et al), resulting from the Gaussian assumption on additive
unobservables. In contrast, in the econometric specification of our paper,
utility functions and distribution of unobserved heterogeneity are left
unspecified, which makes the resulting estimates robust to mis-specification
of unobservables. This generalization makes identification analysis
significantly more challenging, since (a) key steps like Roy's identity and
Slutsky symmetry do not hold in their standard form here, and (b) deriving
the map from \textit{unobserved} heterogeneity in preferences to \textit{%
observed} heterogeneity in demand is not straightforward, and involves
creative use of the single-crossing condition. On the other hand, unlike in
the set-up of Moffitt 1990 or Gan et al 2015, our analysis does not cover
the case of (a) piecewise linear budget constraints, which is already
covered nonparametrically in the above-mentioned papers, and (b) where the
budget constraint is more convex than indifference curves, resulting in
non-unique solutions to a consumer's optimal choice problem.\medskip

\textbf{Plan of the paper}: The rest of the paper is organized as follows.
Section 2 describes the set-up and states the key assumptions. Section 3
presents the analysis of the problem where functional forms of utilities and
how unobserved heterogeneity enter them are not specified. In particular, we
show how to obtain the analogs of Roy's identity and Slutsky symmetry in
this setting, and how to use the resulting system of PDEs to obtain welfare
measures, using data from a large number of markets, each characterized by
its own budget frontier. Section 4 presents an empirical illustration of our
methods. Finally, section 5 concludes. Additional tables and technical
proofs are collected in Appendix 1.

\section{The Set-up and the Problem}

Following Heckman et al \ 2010, denote the key product characteristic by $S$%
, a generic value assumed by $S$ to be $s$, and the hedonic price schedule
describing the relation between price and $S$ by $P\left( S\right) \equiv
P\left( S,\theta \right) $ where $\theta $ is a finite-dimensional parameter
See the introduction for discussions on this point; analogous results when
the budget frontier is left nonparametric are outlined in Appendix 2. We
have data from multiple markets, each with its own $\theta $. For individual
consumers, consumption (of the numeraire) is given by $C=Y-P\left( S,\theta
\right) $ where $Y$ is individual income. Individual preferences are
described by the utility function $U\left( S,C,\eta \right) $ where $\eta $
represents unobserved preference heterogeneity, and $C$ is consumption. In
this, we follow the approach in the recent literature in the econometrics of
hedonic models, cf. Heckman et al 2010 and empirical welfare analysis cf.
Hausman and Newey 2016, Blomquist et al 2021, where the good of interest is
implicitly assumed separable from other goods. So we can view the
consumption decision as the choice over bundles of the good of interest and
the aggregate of all other consumption. For example, in Hausman and Newey
2016, the good of interest is gasoline, in Blomquist et al 2021, it is
after-tax income etc. A household maximizes its utility by choosing $S$
optimally, subject to the budget constraint $Y=P\left( S,\theta \right) +C$.
To keep our focus on identification of welfare effects, we assume for now --
similar to Heckman et al 2010 -- that the function $P\left( S,\theta \right) 
$ in each market is consistently estimable.\footnote{%
In our empirical illustration, we estimate $P\left( S,\theta \right) $ using
instrumental variables that affect the quality of the amenity but are
unlikely to have any direct effect on property rents.}\footnote{%
Indeed, $\theta $ will typically be estimated from the data, but at a
parametric rate, and these estimated $\theta $s will be used subsequently as
regressors, leading to standard measurement error issues. However, variance
of the measurement error in $\theta $ is of order $O\left( n^{-1}\right) $,
where $n$ is the number of observations in each market used to estimate $%
\theta $ in that market. Hence replacing $\theta $ by its estimate will lead
to a very small attenuation bias when $n$ is large. For related discussions,
see Heckman et al, (2010), Section 5.}

\begin{figure}[htbp]
\centering
\includegraphics[width=\textwidth]{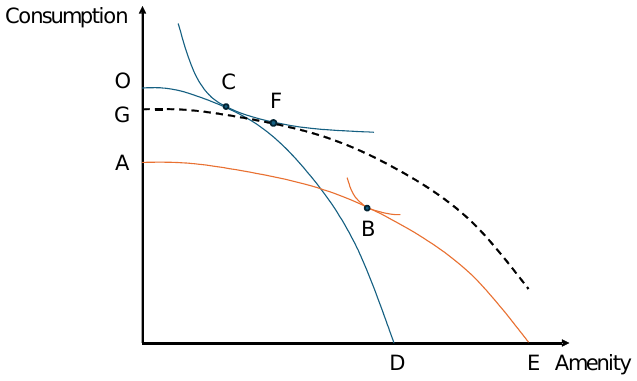}
\caption{Compensating variation when the budget frontier changes.}
\label{fg:compensation_variation}
\end{figure}

In this set-up, we wish to evaluate how a change in the \textit{equilibrium}
hedonic price frontier resulting from a policy intervention affects consumer
welfare. The situation is depicted via Figure \ref{fg:compensation_variation}
where, for ease of exposition, $\eta $ is held fixed.

In Figure \ref{fg:compensation_variation}, consumption is measured on the
vertical axis, and the amenity of interest along the horizontal axis. The
original budget frontier $C=Y-P\left( S,\theta \right) $ is depicted by the
steeper blue curve OD. Utility is maximized at C where the indifference
curve, convex to the origin is tangent to OD. Now suppose that due to a
policy intervention, the equilibrium hedonic price schedule changes, and the
budget frontier shifts to the flatter orange curve AE, whence optimal choice
is B, representing a fall in utility relative to C.

We wish to compute the welfare effect of this change in budget frontier with
the compensating variation, which calculates how much would a household need
to be compensated, so that its maximized utility with the additional income
in the post-intervention situation equals its maximized utility in the
pre-intervention period with the original income. To see this graphically,
consider the dashed curve GF, which is the AE translated vertically up and
is tangent to the original indifference curve at F. Then the compensating
variation, GA$>$0, is the income supplement needed for the
individual facing the blue budget curve so that she can reach utility equal
to what she was enjoying initially.

Given the position of the indifference curves, the CV is positive,
indicating that the consumer is losing as a result of the change, and hence
needs to be compensated by a positive income transfer to restore her utility
to its pre-intervention state. However, if the original indifference curve
were tangent to OD at a point below its intersection with AE, then the shift
of the budget line to AE would lead to a \textit{gain} in utility. Such a
consumer would benefit from the change, and the CV will be negative.
Intuitively speaking, the former type of households value the amenity less
relative to consumption, and so were initially consuming relative lower
quality of amenity. In the new equilibrium following the intervention, costs
rise for areas with lower quality of amenities, and therefore these
households can afford less consumption than before. The latter type of
household values the amenity relatively more, and choose higher quality. The
intervention makes costs lower for areas with good amenities, and hence
expands the budget set of these types of consumers. This reasoning
illustrates that welfare effects of a shift in the budget frontier can be
heterogeneous in both magnitude and sign; hence it is of interest to find
the distribution of welfare as preferences vary across consumers. We now
turn to developing the methods for these calculations.\smallskip

We start our analysis by imposing the following assumptions on the utility
functions. Let $U_{sc}\left( s,c,\eta \right) $, $U_{cc}\left( s,c,\eta
\right) $, etc. denote the second order derivatives of $U$.

\begin{assumption}
(i) $U\left( \cdot ,\cdot ,\eta \right) $ has continuous second-order
derivatives in its first two arguments; (ii) $U\left( \cdot ,\cdot ,\eta
\right) $ is strictly increasing in each argument for any fixed $\eta $;
(iii) $P\left( s,\theta \right) $ is smooth in both $s$ and $\theta $ and is
increasing in $s$ for fixed $\theta $; (iv) for all $s,y,\theta ,\eta $, we
have that%
\begin{equation}
\frac{\partial ^{2}}{\partial s^{2}}\left\{ U\left( s,y-P\left( s;\theta
\right) ,\eta \right) \right\} \equiv \left\{ 
\begin{array}{l}
U_{ss}\left( s,y-P\left( s;\theta \right) ,\eta \right) \\ 
-2\frac{\partial }{\partial s}P\left( s;\theta \right) \times U_{cs}\left(
s,y-P\left( s;\theta \right) ,\eta \right) \\ 
+\left( \frac{\partial P\left( s;\theta \right) }{\partial s}\right)
^{2}\times U_{cc}\left( s,y-P\left( s;\theta \right) ,\eta \right) \\ 
-U_{c}\left( s,y-P\left( s;\theta \right) ,\eta \right) \times \frac{%
\partial ^{2}P\left( s;\theta \right) }{\partial s^{2}}%
\end{array}%
\right\} <0\text{.}  \label{12}
\end{equation}%
\smallskip
\end{assumption}

The smoothness assumption (i) is just a regularity condition enables us to
obtain the key analytical steps for calculating welfare effects using
derivatives; (ii) is non-satiation in $S$ and consumption, which is
intuitive and is a key sufficient condition for our welfare measure, viz.
the compensating variation, to be well-defined. Assumption (iii) says $S$ is
a `desirable' amenity, i.e. consuming more $S$ costs more. Finally,
assumption (iv) is simply the requirement that the function $U\left(
s,y-P\left( s,\theta \right) ,\eta \right) $ is strictly concave in $s$,
which implies that the first order condition yields a global maximum. It is
equivalent to the hedonic budget frontier being `less convex' to the origin
than the indifference curves. In particular, (\ref{12}) holds if the budget
frontier is strictly concave, and utilities are strictly quasiconcave, i.e.,
indifference curves are strictly convex to the origin.

\begin{remark}
Note that, other than smoothness, we have so far made no assumption about
functional form of utilities, how utilities depend on $\eta $ and the
dimension or marginal distribution of $\eta $.
\end{remark}

\section{Results\label{semiparametric}}

In this section, we first derive the analogs of Roy's Identity and Slutsky
symmetry, and then move on to show how to identify welfare effects of a
change in the budget frontier.

\subsection{Roy's Identity and Slutsky Symmetry Under General Heterogeneity}

For ease of exposition, consider the case where $S$ is the only amenity of
interest, the hedonic price function is given by $P\left( S,\theta \right) $%
, where $\theta $ is a vector of parameters. Analogous results when the
budget frontier is left nonparametric appear in the Appendix 2. We will
introduce additional amenities later in Sec 3.3. The utility of an $\eta $%
-type consumer is given by $U\left( s,c,\eta \right) $ where $y$ represents
disposable income, $s$ is the amount of the amenity $S$, $c$ is consumption
of the non-$S$ numeraire, and $\eta $ is unobserved heterogeneity. Denote
the optimal choice by $S^{\ast }\left( y,\theta ,\eta \right) $, which is to
be interpreted as the \textit{structural} demand function.

\begin{remark}
The results of this section concern the object $S^{\ast }\left( y,\theta
,\eta \right) $, which is to be interpreted as the consumption of an $\eta $
type individual when endowed with income $y$. In that sense, it is a
structural demand function. When income is correlated with $\eta $ in a
dataset, recovering the distribution of $S^{\ast }\left( y,\theta ,\eta
\right) $ will require IV or control function type methods, as done in our
empirical illustration below. This does not impact our results in this
section, which concern properties of the structural demand function and its
quantiles.
\end{remark}

Utility maximization and nonsatiation (i.e. Assumption 1(ii) above) imply
that%
\begin{equation}
\left. U_{s}\left( s,y-P\left( s;\theta \right) ,\eta \right) -U_{c}\left(
s,y-P\left( s;\theta \right) ,\eta \right) \frac{\partial }{\partial s}%
P\left( s;\theta \right) \right\vert _{s=S^{\ast }\left( y,\theta ,\eta
\right) }=0  \label{3}
\end{equation}

Finally, the indirect utility function is given by%
\begin{equation}
V\left( y,\theta ,\eta \right) =U\left( S^{\ast }\left( y,\theta ,\eta
\right) ,y-P\left( S^{\ast }\left( y,\theta ,\eta \right) ;\theta \right)
,\eta \right) \text{.}  \label{10}
\end{equation}

Then for fixed $\theta ,\eta $, and given assumption 1(i), we have that $%
V\left( \cdot ,\theta ,\eta \right) $ is differentiable, and the envelope
theorem\ condition holds, i.e.%
\begin{eqnarray}
&&\frac{\partial V\left( y,\theta ,\eta \right) }{\partial y}  \notag \\
&=&\underset{=0\text{, by (\ref{3})}}{\underbrace{\left[ 
\begin{array}{l}
U_{s}\left( S^{\ast }\left( y,\theta ,\eta \right) ,y-P\left( S^{\ast
}\left( y,\theta ,\eta \right) ;\theta \right) \right) \\ 
-U_{c}\left( S^{\ast }\left( y,\theta ,\eta \right) ,y-P\left( S^{\ast
}\left( y,\theta ,\eta \right) ;\theta \right) \right) \times \frac{\partial 
}{\partial s}P\left( S^{\ast }\left( y,\theta ,\eta \right) ;\theta \right)%
\end{array}%
\right] }}\frac{\partial S^{\ast }\left( y,\theta ,\eta \right) }{\partial y}
\notag \\
&&+U_{c}\left( S^{\ast }\left( y,\theta ,\eta \right) ,y-P\left( S^{\ast
}\left( y,\theta ,\eta \right) ;\theta \right) \right)  \notag \\
&=&U_{c}\left( S^{\ast }\left( y,\theta ,\eta \right) ,y-P\left( S^{\ast
}\left( y,\theta ,\eta \right) ;\theta \right) \right)  \label{4}
\end{eqnarray}%
Therefore, $V\left( \cdot ,\theta ,\eta \right) $ is strictly increasing, by
assumption 1(ii).

Further, letting $P_{j}\left( S^{\ast }\left( y,\theta ,\eta \right) ;\theta
\right) =\left. \frac{\partial P\left( s;\theta \right) }{\partial \theta
_{j}}\right\vert _{s=S^{\ast }\left( y,\theta ,\eta \right) }$, we have by
the envelope theorem that%
\begin{equation}
\frac{\partial V\left( y,\theta ,\eta \right) }{\partial \theta _{j}}%
=-U_{c}\left( S^{\ast }\left( y,\theta ,\eta \right) ,y-P\left( S^{\ast
}\left( y,\theta ,\eta \right) ;\theta \right) \right) \times P_{j}\left(
S^{\ast }\left( y,\theta ,\eta \right) ;\theta \right) \text{.}  \label{5a}
\end{equation}

From (\ref{4}) and (\ref{5a}), it follows that for each $j=1,2,...,\dim
\left( \theta \right) $, it must hold that%
\begin{equation}
-\frac{\frac{\partial V\left( y,\theta ,\eta \right) }{\partial \theta _{j}}%
}{\frac{\partial V\left( y,\theta ,\eta \right) }{\partial y}}=P_{j}\left(
S^{\ast }\left( y,\theta ,\eta \right) ;\theta \right)  \label{Roys}
\end{equation}%
which can be interpreted as Roy's identity for a smooth, nonlinear budget
frontier. When the budget constraint is linear, i.e. $P\left( s,\theta
\right) =s^{\prime }\theta $, then (\ref{Roys}) reduces to the standard
Roy's identity%
\begin{equation*}
-\frac{\frac{\partial V\left( y,\theta ,\eta \right) }{\partial \theta _{j}}%
}{\frac{\partial V\left( y,\theta ,\eta \right) }{\partial y}}=S^{\ast
j}\left( y,\theta ,\eta \right) \text{,}
\end{equation*}%
where $S^{\ast j}\left( y,\theta ,\eta \right) $ denotes demand for the $j$%
th amenity at $\left( y,\theta ,\eta \right) $.

Now, suppose we want to measure welfare-effects resulting from a change in $%
\theta $ from $a$ to $b$. A common money-metric measure is the compensating
variation $T\equiv T\left( y,\eta \right) $, which solves%
\begin{equation}
V\left( y+T,b,\eta \right) =V\left( y,a,\eta \right) \text{.}  \label{8}
\end{equation}%
There is a unique solution in $T$, since $\frac{\partial V\left( y,\theta
,\eta \right) }{\partial y}>0$ with probability 1, by (\ref{4}).

To find the distribution of $T$, suppose, initially, that we know the value
of $\eta $, then we can learn $P_{j}\left( s^{\ast }\left( y,\theta ,\eta
\right) ;\theta \right) $ from the hedonic price schedule in the data. Now,
equation (\ref{Roys}) can be rewritten as a system of linear, first-order
partial differential equations of order 1%
\begin{equation}
\frac{\partial V\left( y,\theta ,\eta \right) }{\partial \theta _{j}}+\frac{%
\partial V\left( y,\theta ,\eta \right) }{\partial y}\times P_{j}\left(
S^{\ast }\left( y,\theta ,\eta \right) ;\theta \right) =0\text{.}  \label{9}
\end{equation}%
Therefore, the goal is to solve for (\ref{8}), where $V\left( \cdot \right) $
satisfies (\ref{9}). The key difficulty in calculating welfare effects
nonparametrically from here is that $\eta $ is unobserved. Below, we show
how to address this issue.\smallskip

A second result we will need for welfare to be well-defined is the analog of
classic Slutsky symmetry for nonlinear budget frontiers. This result, like
Roy's identity above, is of independent interest, since it is a fundamental
property of demand, and holds under general heterogeneity. The result is as
follows:

\begin{proposition}
(Slutsky Symmetry) Suppose Assumption 1 is satisfied. Then for all $\eta $,
it holds that%
\begin{eqnarray}
&&\frac{\partial ^{2}P\left( S,\theta \right) }{\partial \theta _{j}\partial
S}\times \left( \frac{\partial S^{\ast }\left( y,\theta ,\eta \right) }{%
\partial \theta _{k}}+\frac{\partial S^{\ast }\left( y,\theta ,\eta \right) 
}{\partial y}\frac{\partial P\left( S^{\ast }\left( y,\theta ,\eta \right)
,\theta \right) }{\partial \theta _{k}}\right)  \notag \\
&=&\frac{\partial ^{2}P\left( S,\theta \right) }{\partial \theta
_{k}\partial S}\times \left( \frac{\partial S^{\ast }\left( y,\theta ,\eta
\right) }{\partial \theta _{j}}+\frac{\partial S^{\ast }\left( y,\theta
,\eta \right) }{\partial y}\frac{\partial P\left( S^{\ast }\left( y,\theta
,\eta \right) ,\theta \right) }{\partial \theta _{j}}\right)  \label{21}
\end{eqnarray}

\begin{proof}
The proof of this result is similar to that of Lemma 1 below, and we omit it
here.
\end{proof}
\end{proposition}

\subsection{Calculation of Money-metric Welfare}

Recall that for calculating welfare effects of changing the budget frontier,
we have to solve for (\ref{8}), where $V\left( \cdot \right) $ satisfies (%
\ref{9}). To address the problem of unobservability of $\eta $, we impose a
single-crossing condition and an assumption of scalar but nonparametric
heterogeneity, and then implement a quantile-based analysis, as follows.

\begin{remark}
Note that none of the results prior to this subsection requires the
restriction on $\eta $ imposed in Assumption 2 below.
\end{remark}

\begin{assumption}
(i) Conditional on observed covariates, the distribution of $\eta $ is
identical across markets, (ii) $\eta $ is a scalar, (iii) the cross-partial
derivatives satisfy $U_{s\eta }\left( s,c,\eta \right) >0$ and $U_{c\eta
}\left( s,c,\eta \right) \leq 0$ for all values of $s,c,\eta $ on the
support of $\left( S,Y-P\left( S,\theta \right) ,\eta \right) $.\smallskip
\end{assumption}

Assumption 2(i) implies that the difference in distribution of demand
(conditional on covariates) across different markets results solely from the
difference in the budget frontiers, i.e. in $\theta $, across markets. This
ultimately enables us to identify the impact of changing $\theta $ via
policy interventions on demand and welfare distributions -- the main goal of
this paper.

\begin{remark}
Assumption 2(i) is the key to identification in the papers cited above,
including Hausman and Newey 2016 and Heckman et al 2010, in that it enables
us to learn about preferences as the budget frontier shifts while the
preference distribution does not.
\end{remark}

In principle, this assumption can be tested in our case by checking if the
distribution of demand for $S$ across individuals with the same income (and
other observed covariates) is identical in different markets with the same
value of $\theta $. The intuitive explanation of why $\theta $ would differ
across markets while the distribution of $\eta $ does not, \textit{%
conditional on observed covariates}, is that the supply side differs across
markets. This is akin to variations in supply identifying demand in standard
simultaneous equation models. While it is conceivable that \textit{within}
geographically large markets, consumers would move according to their
preference for school quality, it is less plausible that consumers would
move from, say, Oxford to Edinburgh or Texas to Maine because schools in
Edinburgh/Maine are better.

Assumption 2(ii) allows for $\eta $ to be a single index of
multi-dimensional underlying heterogeneity, and implies rank invariance,
i.e. the ordering of any two different consumers' demand remains identical
across budget frontiers. Assumption 2(iii) says that the marginal utility
w.r.t. $s$ is strictly increasing, and marginal utility w.r.t. $c$ is
decreasing in $\eta $. Intuitively, this means that higher $\eta $ types
have higher marginal utility w.r.t. $s$ and lower marginal utility w.r.t. $c$%
. This implies the so-called `single crossing' condition, i.e. that the
marginal rate of substitution between $S$ and $C$ is increasing in $\eta $:%
\begin{equation*}
\frac{d}{d\eta }\left( \frac{\frac{\partial U\left( s,c,\eta \right) }{%
\partial s}}{\frac{\partial U\left( s,c,\eta \right) }{\partial c}}\right) >0%
\text{.}
\end{equation*}%
It will be shown below that these assumptions imply that for fixed budget
line, demand for $S$ is strictly monotone in $\eta $. This creates a 1-to-1
map between quantiles of \textit{observed} demand and quantiles of \textit{%
unobserved} preference, which is helpful for identifying welfare. Heckman et
al. (2010) assume, \textit{in addition to 2(i)-(iii),} that utilities are
quasilinear in consumption, so that $U_{c\eta }\left( s,c,\eta \right)
\equiv 0$; assumption 2(iii) is therefore a generalization required to cover
the more general non-quasilinear case.\smallskip

\textbf{Quantile-based Analysis}: Note that if $\eta $ is a scalar and $%
S^{\ast }\left( y,\theta ,\eta \right) $ is strictly monotone and invertible
in $\eta $, then we can interpret the observed $\tau $th quantiles of $%
S^{\ast }\left( y,\theta ,\eta \right) $, conditional on $y$ and $\theta $
as the demand of the individual who is located at the $\tau $th quantile of
the distribution of $\eta $. This is identified by the $\tau $th quantile of
the \textit{structural} demand function across those at income $y$ on the
budget frontier $P\left( s;\theta \right) $, i.e.%
\begin{equation*}
S^{\ast }\left( y,\theta ,F_{\eta }^{-1}\left( \tau \right) \right)
=F_{S^{\ast }\left( y,\theta ,\eta \right) }^{-1}\left( \tau \right) \equiv
q^{\tau }\left( y,\theta \right) \text{,}
\end{equation*}%
where $\tau \in \lbrack 0,1]$, and $q^{\tau }\left( y,\theta \right) \equiv
F_{S^{\ast }\left( y,\theta ,\eta \right) }^{-1}\left( \tau \right) $ equals
the $\tau $th quantile of structural demand for those with income $y$ and
facing a budget frontier characterized by $\theta $.\footnote{%
Hoderlein and Vanhems 2018 consider a similar specification but under linear
budget constraints.} Further, since the indirect utility function%
\begin{equation*}
V\left( y,\theta ,\eta \right) =\max_{s,c}U\left( s,c,\eta \right) \text{
s.t. }c=y-P\left( s,\theta \right) \text{,}
\end{equation*}%
by the envelope theorem, we have that%
\begin{equation}
\frac{\partial }{\partial y}V\left( y,\theta ,\eta \right) =\frac{\partial U%
}{\partial c}\left( S^{\ast }\left( y,\theta ,\eta \right) ,y-P\left(
S^{\ast }\left( y,\theta ,\eta \right) ,\theta \right) \right) >0  \label{15}
\end{equation}%
when utility is strictly increasing in consumption, i.e. assumption 1(ii).

Now, differentiating the LHS of (\ref{3}) w.r.t. $\eta $, we get that

\begin{equation*}
\left( 
\begin{array}{l}
\left\{ 
\begin{array}{l}
U_{ss}\left( s,y-P\left( s;\theta \right) ,\eta \right) \\ 
-2\frac{\partial }{\partial s}P\left( s;\theta \right) \times U_{cs}\left(
s,y-P\left( s;\theta \right) ,\eta \right) \\ 
+\left( \frac{\partial P\left( s;\theta \right) }{\partial s}\right)
^{2}\times U_{cc}\left( s,y-P\left( s;\theta \right) ,\eta \right) \\ 
-U_{c}\left( s,y-P\left( s;\theta \right) ,\eta \right) \times \frac{%
\partial ^{2}P\left( s;\theta \right) }{\partial s^{2}}%
\end{array}%
\right\} \times \frac{dS^{\ast }}{d\eta } \\ 
+U_{s\eta }\left( S^{\ast },y-P\left( S^{\ast };\theta \right) ,\eta \right)
\\ 
-\frac{\partial }{\partial s}P\left( S^{\ast };\theta \right) \times
U_{c\eta }\left( S^{\ast },y-P\left( S^{\ast };\theta \right) ,\eta \right)%
\end{array}%
\right) =0\text{,}
\end{equation*}%
implying%
\begin{equation*}
\frac{dS^{\ast }}{d\eta }=\frac{%
\begin{array}{l}
\frac{\partial }{\partial s}P\left( S^{\ast };\theta \right) \times U_{c\eta
}\left( S^{\ast },y-P\left( S^{\ast };\theta \right) ,\eta \right) \\ 
-U_{s\eta }\left( S^{\ast },y-P\left( S^{\ast };\theta \right) ,\eta \right)%
\end{array}%
}{\left\{ 
\begin{array}{l}
U_{ss}\left( s,y-P\left( s;\theta \right) ,\eta \right) \\ 
-2\frac{\partial }{\partial s}P\left( s;\theta \right) \times U_{cs}\left(
s,y-P\left( s;\theta \right) ,\eta \right) \\ 
+\left( \frac{\partial P\left( s;\theta \right) }{\partial s}\right)
^{2}\times U_{cc}\left( s,y-P\left( s;\theta \right) ,\eta \right) \\ 
-U_{c}\left( s,y-P\left( s;\theta \right) ,\eta \right) \times \frac{%
\partial ^{2}P\left( s;\theta \right) }{\partial s^{2}}%
\end{array}%
\right\} }
\end{equation*}%
The denominator of this expression is strictly negative by (\ref{12}). The
numerator is also strictly negative, by assumptions 1(iii) and 1(iv). Hence $%
\frac{dS^{\ast }}{d\eta }>0$ with probability 1. In Heckman et al. (2010)'s
set-up, utility is quasilinear in consumption, so that $U_{\eta c}\left(
s,c,\eta \right) =0$, which is a special case of our set-up. The
monotonicity of $S^{\ast }$ w.r.t. $\eta $, established above, will be used
below for identifying the distribution of the compensating variation.

We now outline the steps for obtaining welfare effects. Toward that end,
define%
\begin{equation}
Q_{\tau }\left( y,\theta \right) \equiv V\left( y,\theta ,F_{\eta
}^{-1}\left( \tau \right) \right) \text{,}  \label{1}
\end{equation}%
i.e. the indirect utility obtained by an individual with income $y$ and
located at the $\tau $th quantile of unobserved heterogeneity, i.e. whose
value of $\eta $ equals $F_{\eta }^{-1}\left( \tau \right) $, when the price
function is characterized by the parameter $\theta $. First, note from (\ref%
{15}) and (\ref{1}) that%
\begin{equation*}
\frac{\partial }{\partial y}Q_{\tau }\left( y,\theta \right) >0\text{,}
\end{equation*}%
and from (\ref{9}) that for all $\tau \in \left[ 0,1\right] $, and $%
j=1,...,J $,%
\begin{equation}
\frac{\partial Q_{\tau }\left( y,\theta \right) }{\partial \theta _{j}}%
+\left. \frac{\partial P\left( s,\theta \right) }{\partial \theta _{j}}%
\right\vert _{s=q^{\tau }\left( y,\theta \right) }\times \frac{\partial
Q_{\tau }\left( y,\theta \right) }{\partial y}=0\text{.}  \label{6}
\end{equation}

Suppose the parameter $\theta $ characterizing the budget frontier changes
from $a$ to $b$, and we wish to calculate the compensating variation
corresponding to this change for this individual with $\eta =F_{\eta
}^{-1}\left( \tau \right) $. The compensating variation is defined as the
income supplement required to maintain the utility of the consumer, i.e.
solve for $T=T\left( y,\tau \right) $ which satisfies%
\begin{equation}
Q_{\tau }\left( y,a\right) =Q_{\tau }\left( y+T,b\right) \text{, i.e. }%
T\left( y,\tau \right) =Q_{\tau }^{-1}\left( Q_{\tau }\left( y,a\right)
,b\right) -y\text{.}  \label{23}
\end{equation}

\textbf{Slutsky Symmetry}: In order to implement our method of welfare
analysis, it is also necessary to specialize Proposition 1 to quantile
demand.

\begin{lemma}[Slutsky-symmetry for nonlinear budget-sets]
Suppose Assumption 1 and 2 are satisfied, and that the price-amenity
relationship is given by $P\left( s,\theta \right) $ where $\theta $ is of
dimension $d\geq 2$. Let $q^{\tau }\left( y,\theta \right) $ denote the $%
\tau $th quantile of structural demand\ when income is fixed at $y$. Then
for each $j,k\in \left\{ 1,2,...d\right\} $, it holds that%
\begin{equation}
\begin{array}{c}
\frac{\partial ^{2}P\left( q^{\tau }\left( y,\theta \right) ,\theta \right) 
}{\partial \theta _{j}\partial q}\left\{ \frac{\partial q^{\tau }\left(
y,\theta \right) }{\partial y}\frac{\partial P\left( q^{\tau }\left(
y,\theta \right) ,\theta \right) }{\partial \theta _{k}}+\frac{\partial
q^{\tau }\left( y,\theta \right) }{\partial \theta _{k}}\right\} \\ 
=\frac{\partial ^{2}P\left( q^{\tau }\left( y,\theta \right) ,\theta \right) 
}{\partial \theta _{k}\partial q}\left\{ \frac{\partial q^{\tau }\left(
y,\theta \right) }{\partial y}\frac{\partial P\left( q^{\tau }\left(
y,\theta \right) ,\theta \right) }{\partial \theta _{j}}+\frac{\partial
q^{\tau }\left( y,\theta \right) }{\partial \theta _{j}}\right\}%
\end{array}
\label{S6}
\end{equation}%
Proof in Appendix 1.
\end{lemma}

Equation (\ref{S6}) is the analog of ordinary Slutsky symmetry when budget
constraints are smooth and nonlinear. Indeed, in the standard case with
linear budgets, where $P\left( q^{\tau }\left( y,\theta \right) ,\theta
\right) =\theta q^{\tau }\left( y,\theta \right) $, (\ref{S6}) would reduce
to%
\begin{equation*}
\frac{\partial q^{\tau }\left( y,\theta \right) }{\partial y}q^{\tau }\left(
y,\theta \right) +\frac{\partial q^{\tau }\left( y,\theta \right) }{\partial
\theta _{2}}=\frac{\partial q^{\tau }\left( y,\theta \right) }{\partial y}%
q^{\tau }\left( y,\theta \right) +\frac{\partial q^{\tau }\left( y,\theta
\right) }{\partial \theta _{1}}
\end{equation*}%
which is the textbook case of Slutsky symmetry with linear budget frontiers.
Also, note that the proof of the above result does not depend on the number
of goods. \smallskip

The usefulness of the above result is that it makes welfare calculations
path independent. The following theorem states the formal result:

\begin{theorem}
Suppose $\theta \in R^{J}$; the price function $P\left( S,\theta \right) $
and the quantile demand function $q^{\tau }\left( y,\theta \right) $ are
defined on connected open sets, and are twice continuously differentiable on
their domain. Suppose the Slutsky symmetry condition is satisfied, i.e. for
all $j\neq k$ and all values of $\left( y,\theta \right) $ in their support,%
\begin{eqnarray*}
&&\frac{\partial ^{2}P\left( q^{\tau }\left( y,\theta \right) ,\theta
\right) }{\partial \theta _{j}\partial q}\left\{ \frac{\partial q^{\tau
}\left( y,\theta \right) }{\partial y}\frac{\partial P\left( q^{\tau }\left(
y,\theta \right) ,\theta \right) }{\partial \theta _{k}}+\frac{\partial
q^{\tau }\left( y,\theta \right) }{\partial \theta _{k}}\right\} \\
&=&\frac{\partial ^{2}P\left( q^{\tau }\left( y,\theta \right) ,\theta
\right) }{\partial \theta _{k}\partial q}\left\{ \frac{\partial q^{\tau
}\left( y,\theta \right) }{\partial y}\frac{\partial P\left( q^{\tau }\left(
y,\theta \right) ,\theta \right) }{\partial \theta _{j}}+\frac{\partial
q^{\tau }\left( y,\theta \right) }{\partial \theta _{j}}\right\} \text{.}
\end{eqnarray*}%
Then for any path $\theta \left( t\right) $, $t\in \left[ 0,1\right] $, with 
$\theta \left( 0\right) =a$ and $\theta \left( 1\right) =b$, the
compensating variation due to a change in $\theta $ from $a$ to $b$ is given
by $T\left( 1,y,\tau \right) $, where $T\left( t,y\right) $ is the solution
to the ordinary differential equation%
\begin{equation}
\frac{dT\left( t,y,\tau \right) }{dt}=\displaystyle\sum\limits_{j=1}^{J}\frac{d\theta
_{j}\left( t\right) }{dt}\times \frac{\partial P\left( \theta \left(
t\right) ,q^{\tau }\left( \theta \left( t\right) ,y+T\left( t,y,\tau \right)
\right) \right) }{\partial \theta _{j}}\text{,}  \notag
\end{equation}%
with initial condition $T\left( 0,y,\tau \right) =0$, and this solution is
independent of the path $\theta \left( t\right) $.
\end{theorem}

The proof of this result is completely analogous to that of the following
proposition, and uses the fundamental theorem for line integration that line
integrals of gradient fields are path-independent (cf. Courant and John
1989, Sec 1.10, Taylor, 1955 Theorem 13.4.V).

\textbf{Welfare Calculation}: We now outline how to use the above theorem to
calculate welfare effects. In a first step, the function $P\left( s,\theta
\right) $ and the quantile demand $q^{\tau }\left( y,\theta \right) $ are
flexibly estimated from the observed data. A sieve type approximation using,
say, polynomials would be a natural choice. The analytical form of (\ref%
{Roys}) PDEs will change depending on which sieve is used to approximate $%
P\left( s,\theta \right) $ and $q^{\tau }\left( y,\theta \right) $.
Regardless, the linearity of these first-order PDEs in the partial
derivatives of $Q_{\tau }\left( y,\theta \right) $ will be preserved;
consequently, calculating welfare effects would amount to solving the
analogous first-order ordinary differential equations (see the proof of
Proposition 3, below) with polynomials as coefficient functions -- a
standard textbook problem with analytical solutions. The computation steps
are:

\begin{enumerate}
\item Estimate for each market $m$, the hedonic price function $P\left(
s,x,\theta \right) =\sum_{j=1}^{J}\theta _{j}^{\left( m\right) }p_{j}\left(
s,x\right) $, where $X$ denotes a vector of non-$S$ attributes, the $\left\{
p_{j}\left( s,x\right) \right\} _{j=1,...,J}$ denotes a sieve basis, such as
polynomials or splines, the $\theta _{j}^{\left( m\right) }$'s are the
coefficients of the basis terms, $J$ is the number of terms used in the
approximation.

\item Estimate the quantile of log demand in each market as a function of
its own $\theta _{j}$'s and of $y$ by running a quantile regression
(possibly using instruments if $y$ is endogenous) on a sieve of $y$ and the $%
\theta _{j}^{\left( m\right) }$, $j=1,...,J$. Impose the constraints on the
coefficients in Step 2 implied by the Slutsky symmetry eqn. (\ref{S6}).

\item To compute the compensating variation for those at the $\tau $th
quantile of preference due to a change in the value of $\theta $ from $a$ to 
$b$, consider any path $\theta \left( t\right) $, $t\in \left[ 0,1\right] $,
with $\theta \left( 0\right) =a$ and $\theta \left( 1\right) =b$. Then the
compensating variation at income $y$ is given by $T\left( 1,y,\tau \right) $%
, where $T\left( t,y,\tau \right) $ solves the ordinary differential equation%
\begin{equation}
\frac{dT\left( t,y,\tau \right) }{dt}=\displaystyle\sum\limits_{j=1}^{J}\frac{d\theta
_{j}\left( t\right) }{dt}\times \frac{\partial P\left( \theta \left(
t\right) ,q^{\tau }\left( \theta \left( t\right) ,y+T\left( t,y,\tau \right)
\right) \right) }{\partial \theta _{j}}\text{,}  \notag
\end{equation}%
with initial condition $T\left( 0,y,\tau \right) =0$. Under Slutsky
symmetry, the solution is independent of the path $\theta \left( t\right) $.
\end{enumerate}

To illustrate how our method would work in practice, we choose a simple
illustrative specification, viz. that the hedonic price function is
approximated by%
\begin{equation}
P\left( S,\theta \right) =\theta _{1}+\theta _{2}\ln \left( S\right) \text{,}
\label{p}
\end{equation}%
where the values of $\theta _{1},\theta _{2}$ vary across markets. This
corresponds to prices being increasing and concave in the amenity when $%
\theta _{2}>0$.\footnote{%
In our empirical illustration, this specification, after adding other
attributes via their first principal component as described below, gives the
highest adjusted-$R^{2}$ (cf. Pakes 2003, Table 2) among specifications that
include quadratics, and log-log.} Our goal is to find $T$ which solves%
\begin{equation*}
Q_{\tau }(y,b)=Q_{\tau }\left( y+T,a\right) \text{,}
\end{equation*}%
where $Q_{\tau }\left( \cdot ,\theta \right) $ is strictly increasing, and
satisfies the nonlinear analogue of Roy's identity (\ref{Roys})%
\begin{equation}
\left. 
\begin{array}{l}
\frac{\partial Q_{\tau }\left( y,\theta _{1},\theta _{2}\right) }{\partial
\theta _{1}}+\frac{\partial Q_{\tau }\left( y,\theta _{1},\theta _{2}\right) 
}{\partial y}=0\text{,} \\ 
\frac{\partial Q_{\tau }\left( y,\theta _{1},\theta _{2}\right) }{\partial
\theta _{2}}+\ln \left\{ q^{\tau }\left( y,\theta \right) \right\} \times 
\frac{\partial Q_{\tau }\left( y,\theta _{1},\theta _{2}\right) }{\partial y}%
=0\text{.}%
\end{array}%
\right.  \label{7}
\end{equation}%
The term $\ln q^{\tau }\left( y,\theta \right) $ can be identified from the
data by running a quantile regression of (natural log of) the demanded
amenity on individual income (possibly using instruments a la Chernozhukov
and Hansen 2005 when income is endogenous), and the market level $\theta $,
when there are multiple markets, each with its own $\theta $. It is natural
to start with a simple log-linear specification for the quantile regression
function, which may be generalized to a higher order polynomial, spline, etc.%
\begin{equation}
\ln q^{\tau }\left( y,\theta \right) =r_{0}+r_{1}y+r_{2}\theta
_{1}+r_{3}\theta _{2}\text{,}  \label{d}
\end{equation}%
where the $r$ coefficients are obtained via a $\tau $-quantile regression of 
$\ln S$ on individual income and market-level $\theta _{1}$ and $\theta _{2}$%
. Before proceeding further, it is important to verify that (\ref{d}) is
indeed a valid specification for quantile demand, in that it satisfies the
analogue of Slutsky symmetry (\ref{21}). The following proposition states a
necessary condition.

\begin{proposition}
In order for (\ref{d}) to be a valid specification for quantile demand, it
is necessary that $r_{1}+r_{2}=0$. (Proof in Appendix 1).
\end{proposition}

Now suppose the value of the parameter $\theta $ characterizing the price
frontier changes from $a$ to $b$. We wish to find the compensating
variation, which is the hypothetical income transfer that an individual
needs when $\theta =b$ to be able to reach the utility level she had
attained when $\theta $ equalled $a$. Given heterogeneous preferences, the
compensating variation for the same change in $\theta $ is heterogeneous,
and we wish to obtain its distribution. To do this, we fix a value of $\tau
\in \left[ 0,1\right] $, and develop a method to compute the CV for
individuals whose $\eta =F_{\eta }^{-1}\left( \tau \right) $. We then vary $%
\tau $ to generate the CV for different quantiles of $\eta $. The formal
expression for the CV is given by the following proposition.

\begin{proposition}
Suppose the price function $P\left( S,\theta \right) $ and the quantile
demand function $q^{\tau }\left( y,\theta \right) $ are defined on connected
open sets, and are continuously differentiable on their domain. Suppose all
assumptions on preferences stated in assumptions 1 and 2 are satisfied.
Additionally, suppose (\ref{p}) and (\ref{d}) hold with $r_{1}+r_{2}=0$.
Then the compensating variation due to a movement of $\left( \theta
_{1},\theta _{2}\right) $ from $\left( a_{1},a_{2}\right) $ to $\left(
b_{1},b_{2}\right) $ for an individual at $\eta =F_{\eta }^{-1}\left( \tau
\right) $ is independent of the path along which $\left( \theta _{1},\theta
_{2}\right) $ changes, and is given by%
\begin{eqnarray}
T\left( 1,y,\tau \right) &=&e^{r_{1}\left( b_{2}-a_{2}\right) }\left\{ \frac{%
r_{0}}{r_{1}}+y+\frac{a_{2}r_{3}}{r_{1}}+\frac{r_{3}}{r_{1}^{2}}%
-a_{1}\right\}  \notag \\
&&+b_{1}-\frac{r_{0}}{r_{1}}-y-\frac{r_{3}b_{2}}{r_{1}}-\frac{r_{3}}{%
r_{1}^{2}}  \label{cv}
\end{eqnarray}%
Proof in Appendix 1.
\end{proposition}

An analogous exercise can be done for every other quantile, which would
produce different values of the $r$'s in (\ref{d}) and, correspondingly
different values of the compensating variation (\ref{cv}).

\begin{remark}
Note that $Q_{\tau }\left( y,\theta \right) $ defined in (\ref{1}) need 
\textit{not} equal the $\tau $th quantile of the indirect utility $V\left(
y,\theta ,\eta \right) $ because $V\left( y,\theta ,\eta \right) $ need not
be monotonic in $\eta $. Nonetheless, as $\tau $ varies over $\left[ 0,1%
\right] $, we can trace out the distribution of $V\left( y,\theta ,\eta
\right) $. In particular, for each specific quantile, say, $\tau =0.1$ or $%
\tau =0.5$ etc., we get the value of the compensating variation for
individuals with income $y$ and who are at the lowest decile or the median
of unobserved heterogeneity, respectively. These need not equal the the
lowest decile or median respectively of the marginal distribution of the
compensating variation for people with income $y$.
\end{remark}

\subsection{Multiple Attributes\label{Multiple}}

One can incorporate additional hedonic attributes into the above analysis
via an additively separable utility function%
\begin{equation*}
U\left( s,x,c,\eta \right) =U_{1}\left( s,\eta \right) +U_{2}\left( x\right)
+U_{3}\left( c\right) \text{, with }c=y-P\left( s,x;\delta \right) \text{.}
\end{equation*}%
Here $s$ is the key attribute of interest, $x$ represents the other
attributes, $y$ is income, and the hedonic price function is given by $%
P\left( s,x;\delta \right) $, where $\delta =\left( \delta _{1},\ldots
,\delta _{K}\right) ^{\prime }$ is a finite-dimensional vector of
parameters. For computational ease, it may be advisable to use a single
index of the other covariates $x$, e.g. the first principal component of
their covariance.

The utility function is%
\begin{equation*}
V\left( y,\delta ,\eta \right) =\max_{s,x}U\left( s,x,y-P\left( s,x;\delta
\right) ,\eta \right) .
\end{equation*}%
Denote the optimal choices by $S^{\ast }\left( y,\delta ,\eta \right) $ and $%
X^{\ast }\left( y,\delta ,\eta \right) $.

If the attributes are continuous and the utility and price functions are
differentiable, then the first order conditions for an interior maximum are
given by%
\begin{eqnarray}
\frac{\partial U_{1}\left( s,\eta \right) }{\partial s} - U_{3}^{\prime
}\left( y-P\left( s,x;\delta \right) \right) \frac{\partial P\left(
s,x;\delta \right) }{\partial s} &=&0,  \label{multi_foc_s} \\
\nabla _{x}U_{2}\left( x\right) - U_{3}^{\prime }\left( y-P\left( s,x;\delta
\right) \right) \nabla _{x}P\left( s,x;\delta \right) &=&0.
\label{multi_foc_x}
\end{eqnarray}%
We assume that the corresponding Hessian of the objective function with
respect to $\left( s,x\right)$ is negative definite, so that the first order
conditions characterize a unique interior optimum.

We next establish the monotonicity of $S^{\ast }\left( y,\delta ,\eta
\right) $ in unobserved preference heterogeneity $\eta $. Let $H$ denote the
Hessian of the objective function with respect to the choice vector $\left(
s,x\right) $, evaluated at the optimum. Differentiate the system of first
order conditions (\ref{multi_foc_s})--(\ref{multi_foc_x}) with respect to $%
\eta $. Since $\eta $ enters utility only through $U_{1}\left( s,\eta
\right) $, the implicit function theorem gives%
\begin{equation*}
H\left[ 
\begin{array}{c}
\frac{\partial S^{\ast }\left( y,\delta ,\eta \right) }{\partial \eta } \\ 
\frac{\partial X^{\ast }\left( y,\delta ,\eta \right) }{\partial \eta }%
\end{array}%
\right] +\left[ 
\begin{array}{c}
\frac{\partial ^{2}U_{1}\left( S^{\ast },\eta \right) }{\partial s\partial
\eta } \\ 
0%
\end{array}%
\right] =0.
\end{equation*}%
Therefore,%
\begin{equation*}
\left[ 
\begin{array}{c}
\frac{\partial S^{\ast }\left( y,\delta ,\eta \right) }{\partial \eta } \\ 
\frac{\partial X^{\ast }\left( y,\delta ,\eta \right) }{\partial \eta }%
\end{array}%
\right] =-H^{-1}\left[ 
\begin{array}{c}
\frac{\partial ^{2}U_{1}\left( S^{\ast },\eta \right) }{\partial s\partial
\eta } \\ 
0%
\end{array}%
\right] .
\end{equation*}%
Let $H^{11}$ denote the $\left( 1,1\right) $th element of $H^{-1}$. The
first component of the preceding equation is%
\begin{equation}
\frac{\partial S^{\ast }\left( y,\delta ,\eta \right) }{\partial \eta }%
=-H^{11}\frac{\partial ^{2}U_{1}\left( S^{\ast },\eta \right) }{\partial
s\partial \eta }.  \label{multi_monotonicity}
\end{equation}%
Since $H$ is negative definite, $H^{-1}$ is also negative definite and hence 
$H^{11}<0$. Under the single-crossing condition%
\begin{equation*}
\frac{\partial ^{2}U_{1}\left( s,\eta \right) }{\partial s\partial \eta }>0,
\end{equation*}%
it follows from (\ref{multi_monotonicity}) that%
\begin{equation*}
\frac{\partial S^{\ast }\left( y,\delta ,\eta \right) }{\partial \eta }>0.
\end{equation*}%
Thus, conditional on $y$ and $\delta $, the optimal choice $S^{\ast }\left(
y,\delta ,\eta \right) $ is strictly increasing in $\eta $. Consequently,
for each $\tau \in \left[ 0,1\right] $,%
\begin{equation*}
S^{\ast }\left( y,\delta ,F_{\eta }^{-1}\left( \tau \right) \right)
=F_{S^{\ast }\left( y,\delta ,\eta \right) }^{-1}\left( \tau \right) .
\end{equation*}%
Define%
\begin{equation*}
q_{\tau }^{S}\left( y,\delta \right) =S^{\ast }\left( y,\delta ,F_{\eta
}^{-1}\left( \tau \right) \right)
\end{equation*}%
and%
\begin{equation*}
q_{\tau }^{X}\left( y,\delta \right) =X^{\ast }\left( y,\delta ,F_{\eta
}^{-1}\left( \tau \right) \right) .
\end{equation*}%
The latter is the optimal choice of the additional attributes for the same
individual whose unobserved preference heterogeneity equals $F_{\eta
}^{-1}\left( \tau \right) $. It need not equal the marginal $\tau $th
quantile of $X^{\ast }$. Indirect utility is given by%
\begin{equation*}
V\left( y,\delta ,\eta \right) =U_{1}\left( S^{\ast },\eta \right)
+U_{2}\left( X^{\ast }\right) +U_{3}\left( y-P\left( S^{\ast },X^{\ast
};\delta \right) \right) .
\end{equation*}%
By the envelope theorem,%
\begin{equation}
\frac{\partial V\left( y,\delta ,\eta \right) }{\partial y}=U_{3}^{\prime
}\left( y-P\left( S^{\ast },X^{\ast };\delta \right) \right) >0  \label{Vy}
\end{equation}%
and, for each $j=1,\ldots ,K$,%
\begin{equation}
\frac{\partial V\left( y,\delta ,\eta \right) }{\partial \delta _{j}}%
=-U_{3}^{\prime }\left( y-P\left( S^{\ast },X^{\ast };\delta \right) \right) 
\frac{\partial P\left( S^{\ast },X^{\ast };\delta \right) }{\partial \delta
_{j}}.  \label{multi_Vdelta}
\end{equation}%
Combining the two previous displays gives the analogue of Roy's identity for
the multiple-attribute case,%
\begin{equation}
\frac{\partial V\left( y,\delta ,\eta \right) }{\partial \delta _{j}}+\frac{%
\partial V\left( y,\delta ,\eta \right) }{\partial y}\frac{\partial P\left(
S^{\ast },X^{\ast };\delta \right) }{\partial \delta _{j}}=0,\qquad
j=1,\ldots ,K.  \label{multi_roy}
\end{equation}

Now define%
\begin{equation*}
Q_{\tau }\left( y,\delta \right) =V\left( y,\delta ,F_{\eta }^{-1}\left(
\tau \right) \right) .
\end{equation*}%
Evaluating (\ref{multi_roy}) at $\eta =F_{\eta }^{-1}\left( \tau \right) $
gives%
\begin{equation}
\frac{\partial Q_{\tau }\left( y,\delta \right) }{\partial \delta _{j}}+%
\frac{\partial Q_{\tau }\left( y,\delta \right) }{\partial y}\frac{\partial
P\left( q_{\tau }^{S}\left( y,\delta \right) ,q_{\tau }^{X}\left( y,\delta
\right) ;\delta \right) }{\partial \delta _{j}}=0,\qquad j=1,\ldots ,K.
\label{multi_quantile_roy}
\end{equation}

Suppose a policy intervention changes the equilibrium hedonic price surface
from one characterized by $\delta =a$ to one characterized by $\delta =b$.
For an individual at $\eta =F_{\eta }^{-1}\left( \tau \right) $, define
compensating variation $T\left( y,\tau \right) $ by%
\begin{equation}
Q_{\tau }\left( y+T\left( y,\tau \right) ,b\right) =Q_{\tau }\left(
y,a\right) .  \label{multi_cv_def}
\end{equation}

Consider a continuously differentiable path $\delta \left( t\right) $, $t\in %
\left[ 0,1\right] $, with $\delta \left( 0\right) =a$ and $\delta \left(
1\right) =b$. Let $T\left( t;y,\tau \right) $ denote compensating variation
accumulated along this path, with $T\left( 0;y,\tau \right) =0$. Holding
utility constant along the path gives%
\begin{equation*}
Q_{\tau }\left( y+T\left( t;y,\tau \right) ,\delta \left( t\right) \right)
=Q_{\tau }\left( y,a\right) .
\end{equation*}%
Differentiating with respect to $t$ yields%
\begin{equation*}
\frac{\partial Q_{\tau }}{\partial y}\frac{dT\left( t;y,\tau \right) }{dt}%
+\sum_{j=1}^{K}\frac{\partial Q_{\tau }}{\partial \delta _{j}}\frac{d\delta
_{j}\left( t\right) }{dt}=0.
\end{equation*}%
Using (\ref{multi_quantile_roy}), we obtain%
\begin{eqnarray}
\frac{dT\left( t;y,\tau \right) }{dt} &=&\sum_{j=1}^{K}\frac{d\delta
_{j}\left( t\right) }{dt}\frac{\partial P\left( q_{\tau }^{S}\left(
y+T\left( t;y,\tau \right) ,\delta \left( t\right) \right) ,q_{\tau
}^{X}\left( y+T\left( t;y,\tau \right) ,\delta \left( t\right) \right)
;\delta \left( t\right) \right) }{\partial \delta _{j}},
\label{multi_cv_ode} \\
T\left( 0;y,\tau \right) &=&0.  \notag
\end{eqnarray}

Accordingly, once the hedonic price function $P\left( s,x;\delta \right) $
and the relevant structural quantile demand functions have been estimated,
compensating variation can be calculated by solving (\ref{multi_cv_ode})
along a path connecting the pre- and post-intervention values of $\delta $.
Subject to the corresponding integrability restrictions, the resulting
compensating variation is independent of the particular path used to connect
the initial and terminal price surfaces.

\section{Empirical Illustration}

In this section, we provide a simple empirical illustration of how our
theoretical methods are to be applied in practice. For this exercise, we use
the motivating example mentioned in the introduction, viz. the relation
between house rental price and neighbourhood school quality. We emphasize
that the nature of this illustration is essentially a `proof of concept',
rather than a full blown empirical analysis of the substantive problem.

\subsection{Data\label{Data}}

The dataset used for our illustration comes from the restricted-access
version of Wave 2015 of English Housing Survey (DCLG{\normalsize , 2018}),
which is a nationally representative survey on housing stock, conditions,
and household characteristics.\footnote{%
An alternative data source would be the UK Household Longitudinal Study
(UKHLS), which provides longitudinal data on households and housing
circumstances. However, UKHLS lacks property characteristics such as floor
area, detailed property type and property age, which are important for
modeling housing prices. Moreover, UKHLS is a household panel survey where
the sample is not designed to be representative of the housing stock. I The
English Housing Survey (EHS) is designed to produce a representative sample
of dwellings, making it more suitable for our purposes.} We use data on
rents to avoid measurement error in occupier-reported property prices.%
\footnote{%
Earlier waves of the EHS, which included both owner-reported valuations and
Valuation Office Agency (VOA) assessments, show that owner-occupiers tend to
overestimate the value of their homes, with only 40\% of household
valuations falling within 10\% of the VOA value. Mis-valuation is
systematically correlated with owner and property characteristics such as
income, property value, length of tenure, age, and household composition
(DLUHC, 2022). VOA assessments have been stopped in 2010 and are not
available in our wave.}

For each property, we observe the annual rent as well as a range of
structural features of the property, including floor area, number of floors,
dwelling type (terrace, detached, flat, etc.), age, number of bathrooms,
bedrooms, and living rooms. In addition, we observe an index of local level
of multiple deprivation, measured at the level of the so-called `Lower Layer
Super Output Area', each containing between 400 to 1,200 households. We also
observe a set of household characteristics, including net annual income of
the household, whether the household receives housing benefits, and tenure
type, i.e. whether renting privately, via local authority provision, or
through housing associations. We cannot infer the exact location of the
closest commercial centre and therefore its distance from the property, nor
its public transport access -- two potentially important determinants of
home value.

To proxy for school quality, we use the average point score per pupil for
secondary schools. The point score comes from the Key Stage 4 data in the
School Performance Tables, commonly known as \textit{league tables}.%
\footnote{%
Key Stage 4 represents the two years of education for students aged 14 to
16, corresponding to Years 10 and 11 in the English education system.} These
data are publicly available via the UK government's official website,
GOV.UK. We exclude independent, i.e. private, schools and schools for
children with special educational needs, i.e. special schools, because they
follow different admission procedures and cater to a distinct population.

Each property is matched to the nearest school based on postcode proximity.
The matching process was carried out using the open-source geographic
information system software, QGIS. Distance between home and school is
important for school allocation in England. For oversubscribed schools,
local authorities typically operate a centralized admissions system that
prioritizes applicants based on published criteria, e.g. straight-line
distance.

In our dataset, we exclude 110 cases where household incomes are negative
after accounting for rent. The final sample includes 6,500 properties, each
matched to the nearest school. Table \ref{tab:desc} of Appendix 1 presents
the descriptive statistics for the dataset. On average, households in our
sample have a post-tax weekly income of \pounds 416, with around \pounds 120
allocated to rent. Around 32 percent of households rent privately, while the
remainder are social renters, either from local authorities or housing
associations. Approximately 54 percent of the respondents receive housing
benefits, and 54 percent reside in areas within the three most deprived
deciles. For the purpose of empirical application, we reduce the
dimensionality of the property and household characteristics by using its
first principal component.\footnote{%
Appendix Table \ref{tab:pc_cor} reports the correlations between the
original variables, while Appendix Table \ref{tab:pc_loadings} provides the
loadings on the first principal component.}

Assumption 1(iii) requires the price schedule $P\left( s,\theta \right) $ to
be smooth in $s$ within each market, which is plausible only if school
quality varies densely within markets rather than taking a handful of
values. Table \ref{tab:supp} of Appendix 1 reports, for each of the nine
markets -- the official regions of England, see Step 2 of Section \ref%
{Calculation_nonp} -- the number of distinct values of school quality
observed in the estimation sample, together with the number of distinct
schools matched to properties in that market. The support is rich in every
market, ranging from 118 distinct quality levels in the North East to 340 in
the South East, with each market containing between 447 and 934 properties.
School quality is therefore observed at many closely spaced values within a
market rather than at a small number of isolated points, so the
within-market price schedule is defined over an effectively continuous
support.

Table \ref{tab:main_res} presents the results of the hedonic regression for
property rental prices for the whole of England, where Column 1 presents the
OLS results and Column 2 presents the estimates from the IV approach.

To address potential endogeneity of school quality in the regression of
rental prices on school quality, we exploit variation introduced by the
Academies Act 2010, a major education reform in England. Under this reform,
publicly funded schools could apply to convert to `academy' status, thereby
gaining greater autonomy from local government control and increased
discretion over budgets, staffing, curriculum, and governance. Existing
evidence suggests that conversion is positively associated with higher
school performance, although the effect varies depending on the number of
years since conversion (Andrews et al., 2017). This policy reform was also
used as a source of exogenous variation by Duchini et al 2020 to study the
effect of education on youth crime. To capture this variation, we use both
an indicator for academy status and the number of years since conversion as
instruments for school quality.

Typical approaches to address endogeneity in similar contexts include
exploiting school district boundaries (see e.g. Black 1999) or using
variation induced by natural experiments arising from policy changes
(Hussain, 2023; Machin and Salvanes, 2016). We cannot adopt the second
approach, as there are generally no school districts in the English setting,
apart from a few areas that do not provide large enough samples for our type
of analysis. Our approach is closer in spirit to that of Hussain 2023, who
also studies England and relies on school inspections that occur
quasi-randomly once every few years as a source of exogenous variation. We,
however, cannot use the same source of variation, as the timing of
inspections has been linked to school performance since 2008 and, hence,
stopped being quasi-random.

A key concern in using conversion to academy as an instrument is whether it
is correlated with unobserved local housing market determinants beyond those
controlled for in the model. We show that the correlations between local
household income and academy status and years since conversion are modest,
equalling 0.05 and 0.06, respectively, and correlations with household
savings are similarly small (0.03 and 0.03). The corresponding regression
coefficients, controlling for other included covariates yield highly
insignificant t-statistics. While local deprivation is moderately correlated
with conversion status (0.24) and years since conversion (0.25), local
deprivation is explicitly controlled for in the regression.

The results in Table \ref{tab:main_res} show that the rents are positively
and significantly associated with the quality of the nearest school: an
increase in one standard deviation in school quality is associated with
extra \pounds 12.5 or 10 percent of weekly rent in the IV specification,
which we use as the baseline.\footnote{%
One standard deviation of logarithm of total average point score per pupil
equals 0.20.} The literature on the relationships between school quality and
rental prices is scarce, making it hard to compare our result to earlier
findings.\footnote{%
To the best of our knowledge, Bayer, Ferreira, and McMillan (2007) is the
only paper that includes an analysis of rental prices. They use a dataset
that combines rental and purchase prices to study the association with
elementary school quality. They found that households are willing to pay
less than 1 percent more in house prices when the average school performance
increases by 5 percent. We estimate a higher relationships of 2.6 percent,
focusing solely on secondary schools and renters.} Our estimate is at the
upper bound of what is typically found in the larger literature that focuses
on purchase property prices (Machin, 2011).

\subsubsection{Computation Steps\label{Calculation_nonp}}

The computation of welfare is done through the following steps, where for
simplicity, we use $\tau =0.5$ for illustration.

\begin{enumerate}
\item Construct the scalar index $X$ which equals the first principal
component of all non-$S$ attributes (STATA command pca). This is done to
reduce the dimension of the covariates.

\item Divide properties into $M$ markets. For each market, estimate the
price function by regressing price of unit on $\ln S$ and $X$\footnote{%
Following Hussain (2023), who finds a convex relationship between school
quality and housing prices, we explore a quadratic specification for school
quality. In our data, however, the estimated relationship is concave, so we
proceed with a logarithmic specification instead.} using the IV described
above; call the coefficients $\theta _{1m},\theta _{2m},\delta _{m}$, $%
m=1,...M$%
\begin{equation}
P_{mi}\left( S,X\right) =\theta _{1m\left( i\right) }+\theta _{2m\left(
i\right) }\ln S_{i}+\delta _{m\left( i\right) }X_{m\left( i\right) }
\label{19}
\end{equation}%
where $m\left( i\right) $ is the market in which $i$ lives. Depending on the
context, these markets may be defined purely geographically, e.g. each
county or each postcode being one market; alternatively, they may be defined
by clustering the coefficients of the above regression, i.e. locations with
coefficients in the same cluster constituting one market. In our
illustration below, we choose each official region of England as one
distinct market. Equation (\ref{19}) is a simplification for aiding
computation which we use because our main purpose here is to illustrate how
to apply our methods. We can, in principle, consider a more involved
functional form such as a sieve with B-spilnes, orthogonal polynomials etc.

\item Run a linear median regression (qreg in STATA), using all
observations, of $\ln S_{i}$ on an intercept and $y_{i},\theta _{1m\left(
i\right) },\theta _{2m\left( i\right) }$ and $\delta _{m\left( i\right) }$%
\begin{equation}
med\left( \ln S_{i}|y_{i},\theta _{1m\left( i\right) },\theta _{2m\left(
i\right) },\delta _{m\left( i\right) }\right) =r_{0}+\underset{\text{%
imposing }r_{1}+r_{2}=0}{\underbrace{r_{1}\left( y_{i}-\theta _{1m\left(
i\right) }\right) }}+r_{3}\theta _{2m\left( i\right) }+r_{4}\delta _{m\left(
i\right) }\text{.}  \label{20}
\end{equation}

\item Fix a value of $y=y_{0}$, $\delta =\delta _{0}$ (say, median values of 
$y$ and $\delta $ in the data).

\item Then consider the change in $\theta _{1},\theta _{2}$ from $\left(
a_{1},a_{2}\right) $ to $\left( b_{1},b_{2}\right) $ (say, from the bottom
quartile to top quartile).

\item Calculate compensating variation as%
\begin{eqnarray*}
T &=&e^{r_{1}\left( b_{2}-a_{2}\right) }\left\{ \frac{r_{0}}{r_{1}}+y_{0}+%
\frac{a_{2}r_{3}}{r_{1}}+\frac{r_{3}}{r_{1}^{2}}-a_{1}\right\} \\
&&+b_{1}-\frac{r_{0}}{r_{1}}-y_{0}-\frac{r_{3}b_{2}}{r_{1}}-\frac{r_{3}}{%
r_{1}^{2}}\text{.}
\end{eqnarray*}

\item Replace median in Step 3 to other quantiles, e.g. $\tau =0.25,0.75$
etc. and repeat steps 4-6.
\end{enumerate}

\subsubsection{Endogeneity Issues\label{Endog}}

Our theoretical results above concern structural objects, viz. the price
equation $P\left( S,\theta \right) $ and quantile demand $q^{\tau }\left(
y,\theta \right) $. In order to implement our results empirically, we need
to address standard endogeneity issues, under which the sample demand
quantiles do not correspond to the structural objects of interest. There are
three potential endogeneity issues here. The first is the endogeneity of the
quality of amenity in equation (\ref{19}). This can be addressed either
using instrumental variables, such as policies that affect provision of the
amenity but have no direct impact on housing costs, or by using border
discontinuity type strategies, such as those in Black, 1999. For the purpose
of our empirical illustration below, we follow Hausman and Newey 2016, and
assume that conditional on observables, budgets sets and preferences are
uncorrelated. This is both for (A) reasons of simplicity, given that we are
not contributing to solving the omitted characteristics problem that is well
studied in the literature, and (B) we do not have access to standard
instruments (like inherited income, border discontinuities etc.). The second
source of endogeneity is random measurement error in reported income in
equation (\ref{20}). We use savings -- elicited via an independent question
about the amount of annual savings -- as an instrument here as it is
correlated with true income and is likely uncorrelated with random
measurement error in reported income. The third is potential endogeneity of
variables like $\theta _{2m\left( i\right) }$ in (\ref{20}) due to prior
sorting of the population into regions. Assumption 2(i) requires the
distribution of $\eta $ to be the same in every market once observed
covariates are conditioned on, while our identification rests on
cross-market variation in $\theta $. The two are mutually consistent when
the differences between markets arise from the supply side. English regions
differ in the elasticity of housing supply, in land and planning
constraints, in the age and composition of the housing stock, and in how
school quality is spread across space. Each of these shifts the rent
frontier that households face without shifting the distribution of tastes
among renters. We view our welfare analysis both as conditional on the prior
sorting decision, and also as a short-run exercise, in the sense that a
policy intervention changing the values of the $\theta $'s in (\ref{19})
may, in the long run, cause re-sorting of the population and also change the
values of $r$ in (\ref{20}), which we do not address here and leave for
future research. Lastly, because our method is based on data from many
markets, we avoid the pitfall (Brown and Rosen 1982) of spurious hedonic
regression of demand on price using data from a single market that
essentially returns the hedonic price frontier rather than the demand curve.

\subsection{Empirical Results}

In this section, we report the results obtained by applying the methods
outlined in Section \ref{Calculation_nonp} with a single amenity, viz.
school-quality. For the estimation we split the dataset into 9 markets,
represented by English regions: North East, North West, Yorkshire and the
Humber, East Midlands, West Midlands, East of England, London, South East,
and South West. Table \ref{tab:reg_by_region} presents the results of
hedonic regressions for rental prices estimated for different regions. The
model is described in Step 2 of Section \ref{Calculation_nonp}. School
quality is instrumented with school's academy status and the duration of
this status.\footnote{%
Out of 6,500 households in the sample, 35.83 percent or 2,329 households
live near a converted school. Out of converted schools 153 converted last
year, 330 converted 2 years ago, 681 converted 3 years ago, and 1,165
converted 4 years ago or earlier.} None of the F-statistics for regional
regressions is below 15. The estimated gradient is positive in eight of the
nine regions, and statistically significant in three of them: the North
West, the East of England and London. Our identification requires that these
gradients differ across markets, since it is the variation in $\theta _{2m}$
that identifies demand. A Wald test computed from a pooled regression
interacting every term with region rejects their equality (${\chi}%
^{2}(8)=57.26$, p$<$0.001).

Treating regions as markets requires that households choose among schools
within a region rather than across regions. The distribution of school
quality supports this. A regression of school quality on region indicators
gives an $R^{2}$ of 0.045, so about 95 per cent of the variation in school
quality lies within regions rather than between them. The moments at the
foot of Table \ref{tab:reg_by_region} show the same market by market: mean
school quality ranges from 341 in the East Midlands to 387 in London, a span
of 14 per cent of its average across markets, while the standard deviation
within a region is around 65 points. A household that is interested in a
particular level of school quality therefore has meaningful choice inside
its own region.

Table \ref{tab:q_iv} reports the results of a linear quantile regression for
the logarithm of the school quality presented in Step 3 of Section \ref%
{Calculation_nonp}. The estimate of $r_{1}$ is positive and statistically
significant across quartiles: increase in income positively affects the
demand for school quality across the distribution of preferences for school
quality. The estimate of $r_{3}$ is negative across quartiles: when the
relationships between school quality and prices grow stronger, demand for
school decreases. The latter coefficients are significant for the 50th and
75th percentile, but not for the 25th percentile of the distribution of
school quality.

\begin{figure}[htbp]
\centering
\includegraphics[width=4.2731in]{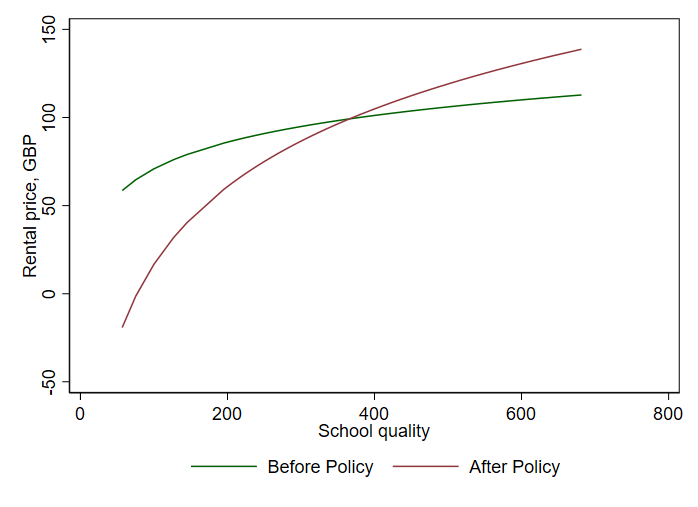}
\caption{Change in the relationships between rents and school quality as a result of policy change}
\label{fig:np_policy_change}
\end{figure}

We now introduce a hypothetical policy change that increases the sensitivity
of rental prices to school quality. An example of such policy would be
introducing more strict distance criteria for school admission. In
particular, we change the relationships between the school quality and
rental prices, $\theta _{2m(i)}$, from the 25th percentile of the
distribution across the markets (21.35 in Yorkshire and the Humber) to the
75th percentile (63.38 in the North West), and analogously exchange the
constant term. By design, rental prices near better schools should increase,
while the prices near worse schools should go down.

Figure \ref{fig:np_policy_change} presents the change in the relationship
between school quality and rental payments as a result of the policy change.
In our sample, the rental price frontier rotates approximately around the
median of the school quality distribution (360.00, which is close to the
mean of 363.97), meaning that rents increase for roughly half of the sample
-- those located near the highest-quality schools -- and decrease for
households near lower-performing schools.

\begin{figure}[htbp]
\centering
\includegraphics[width=4.4865in]{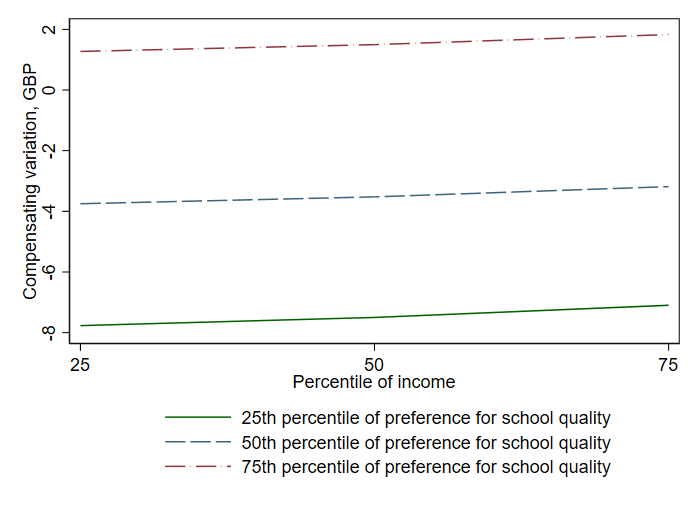}
\caption{Compensating variation estimates for quartiles of preference for school quality, GBP.}
\label{fig:np_cv}
\end{figure}

The resulting estimates of compensating variation for different quantiles of 
$\eta $ evaluated at specific quantiles of income are presented in Table \ref%
{tab:cv_nonp1} and Figure \ref{fig:np_cv}. The policy change results in
welfare loss for households near the highest performing schools, and a
welfare gain for those near the schools at 50th percentile and below.
Compensating variation at the median preference for school quality ($\eta
=F_{\eta }^{-1}\left( 0.5\right) $), evaluated at $y_{0}$ equal to the
median income,\ equals -\pounds 3.5,\ which constitutes 3\ percent of
average weekly rent in England. Households at the same income that live near
schools of higher quality (i.e. have higher $\eta $) experience a lower
welfare gain (or greater welfare loss) than those living near worse schools.
This is to be expected, as their rents increase more, see Figure \ref%
{fig:np_policy_change}. For households at identical percentiles of $\eta $,
those with higher incomes experience lower gain (or higher loss). This is
again to be expected, as comparatively richer households are likely to live
near better schools, where housing costs rise after the policy change,
leaving lesser funds for consumption. To understand why the welfare curves
in Fig. \ref{fig:np_cv} at different percentiles of preference distributions
are nearly parallel, i.e. $\frac{\partial }{\partial \tau }\frac{\partial
\left( T\right) }{\partial y}\simeq 0$, recall that the CV at the $\tau $th
quantile of preferences is given by%
\begin{eqnarray*}
T\left( 1,y,\tau \right) &=&e^{r_{1}^{\tau }\left( b_{2}-a_{2}\right)
}\left\{ -a_{1}+\frac{r_{0}^{\tau }}{r_{1}^{\tau }}+y+\frac{a_{2}r_{3}^{\tau
}}{r_{1}^{\tau }}+\frac{r_{3}^{\tau }}{r_{1}^{\tau 2}}\right\} \\
&&+b_{1}-\frac{r_{0}^{\tau }}{r_{1}^{\tau }}-y-\frac{r_{3}^{\tau }b_{2}}{%
r_{1}^{\tau }}-\frac{r_{3}^{\tau }}{r_{1}^{\tau 2}}\text{.}
\end{eqnarray*}%
Therefore, $\frac{\partial }{\partial \tau }\frac{\partial \left( T\right) }{%
\partial y}=\frac{\partial }{\partial \tau }e\left( ^{r_{1}^{\tau }\left(
b_{2}-a_{2}\right) }\right) \simeq 0$ when $\frac{\partial r_{1}^{\tau }}{%
\partial \tau }\simeq 0$, i.e. different quantiles of demand are equally
sensitive to income. This is corroborated by Table 8 in Appendix 1, where
the coefficient on $\left( y-\theta _{1}\right) $ is nearly identical across
quantiles.

As a further robustness check, we re-estimate the demand equation of Step 3
on two subsamples of renters. About a third of our sample rents privately.
The remainder are social tenants, whose rents are set administratively and
whose properties are allocated by need rather than chosen. We retain all
renters in our main specification, which makes our estimates conservative.
In both subsamples we restrict the sample at Step 3 only, and estimate the
hedonic frontiers of Step 2 on the full regional samples throughout. The
frontier is a market-level object: it is the schedule of rents faced by
every household transacting in the market, and it is priced by all of those
transactions, not only by the households whose demand we go on to study.
Demand is a household-level object, so a restriction on households belongs
at Step 3. Estimating it this way also holds the number of markets fixed at
nine across the three samples, which matters because $r_{3}$ and $r_{4}$ are
identified from variation across markets rather than within them.

Panel B of Table \ref{tab:q_iv_subsamples} restricts the sample to the 2,063
private renters. The estimate of $r_{3}$ is negative and statistically
significant at all three quartiles, and larger in absolute value than in the
full sample. In the full sample, it is not significant at the 25th
percentile. The estimate of $r_{1}$ is nearly unchanged and remains flat
across quartiles. Panel C restricts the sample further to the 947 private
renters with children. These are the households for whom school quality is
most likely to matter. These estimates are less precise, as one would expect
from a sample this size spread across nine markets, but their signs and
magnitudes are unchanged. The compensating variation estimates in Table \ref%
{tab:cv_subsamples} follow the pattern of the main sample in both
subsamples: households at the lower quartiles of preference for school
quality gain, those at the 75th percentile lose, and at identical
percentiles of $\eta $ those with higher incomes experience lower gain or
higher loss.

As a robustness check, we address measurement error in reported income in
the estimation of the demand for schools in Step 3 of Section \ref%
{Calculation_nonp}. The term $y-\theta _{1m(i)}$ is instrumented with $%
s-\theta _{1m(i)}$, where $s$ is the level of past year savings reported by
the household, using the `ivqreg' command in STATA, implementing the method
of Chernozhukov and Hansen (2005). Savings are informative about a
household's true income and are unlikely to affect schooling choice other
than through it. Whether their reporting error is unrelated to the error in
reported income is less clear, since the two are collected in the same
interview, so we treat the instrumented estimates as a diagnostic rather
than as our preferred specification. Appendix Tables \ref{tab:q_2iv} and \ref%
{tab:cv_nonp1_2iv} present the results, which stay qualitatively similar.
The estimate of $r_{1}$ is roughly twice as large under IV at every
quartile. This is what classical measurement error predicts, since
attenuation would bias $r_{1}$ towards zero, and it is the opposite of what
correlation between income and the preference for schooling would produce.
The two are not mutually exclusive, so we read this as suggestive rather
than conclusive.

Our framework assumes that households can adjust their location without
cost, as static models of residential sorting do (see e.g. Bayer et al
2007). Bayer et al (2016) relax this assumption, treating the financial cost
of moving as six per cent of house value and estimating the psychological
cost of moving to be large. The compensating variations we report are the
compensation a household requires when it can reoptimize freely. If
consuming a different amount of school quality requires a costly move, the
compensation it requires is larger, so our figures overstate the gains and
understate the losses. The discrepancy is confined to households that would
move.

Our estimates also describe renters alone. We use rent because it is the
price a household actually pays over a period, and because owner-reported
property values carry measurement error that is correlated with the
characteristics of the owner, as Section \ref{Data} explains. Renters differ
from owner-occupiers in income, age and length of residence, so the
compensating variations above should be read as describing renters rather
than English households as a whole. Our estimates also should be treated as
short-run figures, as noted in Section \ref{Endog}: each one values a change
in the rent frontier holding the household's market fixed, and none of them
is a prediction of how households would re-sort across markets if the price
function were different.

\section{Conclusion}

In this paper, we analyze demand for attributes in markets characterized by
heterogeneous consumers facing smooth, nonlinear budget constraints. We
provide an econometric method of computing welfare effects of changes in the
equilibrium budget frontier. The two building blocks of our result are (a)
an analog of Roy's identity, and (b) an analog of Slutsky symmetry for
structural demand, which hold under unrestricted heterogeneity. These are
two new theoretical results of independent interest and provide the
fundamental link between what is observed (i.e. demand) and what is sought
(preferences and welfare). Indeed, for the canonical case of linear budget
frontiers, these two results appear in every undergraduate textbook on
microeconomics; however, their analogues for nonlinear budgets were
previously unknown. Furthermore, we establish here that the above results
hold no matter what the dimension/distribution of unobserved preference
heterogeneity is and irrespective of whether it is known or unknown. Next,
we show how dimension restrictions on unobserved heterogeneity and a
single-crossing property of preferences enable one to identify the
coefficient functions in the system of PDEs generated by Roy's identity, and
then use Slutsky symmetry to establish uniqueness of welfare effects
resulting from a change in the budget frontier. We provide a simple
numerical illustration of our methods by calculating welfare effects of a
hypothetical change in the equilibrium relationship between property rent
and neighboring school quality, using survey data from England.

There is a plethora of practical situations where consumers choose an
alternative among a large available set but based on a small number of key
attributes. Such markets are conveniently modelled via hedonics. In these
markets, changes in market conditions, such as deliberate policy
interventions or unanticipated market shocks can change the equilibrium
relationship between price of a product and its constituent attributes. The
methods developed in the present paper are useful for evaluating the welfare
effect of such changes. As such, they are applicable to a wide variety of
practical settings involving changes in technology, government policy
regarding taxes/subsidies, provision of public goods like healthcare,
schools and so forth. For example, a switch in production technology
replacing low-skill human jobs with AI would make the equilibrium wage
function more sensitive to skill-levels; increased competition in engine
manufacturing would make equilibrium car prices less dependent on fuel
efficiency and more on sleekness of design etc. These would constitute a
change in the equilibrium hedonic budget frontier, whose welfare
consequences can be inferred using the methods outlined above.

In our analysis so far, we have specified the hedonic price function
parametrically. In principle, it would be worthwhile to extend this to a
nonparametric form and to derive the functional versions of Roy's identity
and Slutsky symmetry using functional derivatives. We leave this task to
future research.\newpage

\onehalfspacing

\section*{Tables}

\begin{center}
\begin{table}[h]
\caption{Hedonic regression results for England.}
\label{tab:main_res}\centering%
\begin{tabular}{lcc}
\hline
& Weekly rent & Weekly rent \\ 
& OLS & IV \\ \hline
Logarithm of total average point score per pupil & 52.604*** & 63.139*** \\ 
& (3.621) & (10.026) \\ 
First principal component of property characteristics & 8.686*** & 8.683***
\\ 
& (0.365) & (0.365) \\ 
Constant & -225.788*** & -287.712*** \\ 
& (21.350) & (58.957) \\ \hline
R-squared & 0.11 & 0.11 \\ 
Number of observations & 6500 & 6500 \\ \hline
\end{tabular}
\newline
\end{table}

{\footnotesize \textit{Note}: To control for endogeneity, the logarithm of
total average point score per pupil in the IV model is instrumented with
school's academy status and the number of years since conversion.
F-statistics for the first stage equals 487.7. Standard errors in
parentheses. * p$<$0.10, **p$<$0.05, *** p$<$0.01}

\newpage

\bigskip {\footnotesize 
\begin{landscape}

\begin{table}[h]
\centering
\footnotesize
\caption{Hedonic regression results by region.}
\label{tab:reg_by_region}
\begin{tabular}{lccccccccc}
\hline
                            & North East & North West & Yorkshire and & E. Midlands & W. Midlands & East      & London   & South East & South West \\ 
                            & & &  the Humber &  &  &      &  &  & \\ \hline
Logarithm of school quality & 13.958      & 63.382***  & 21.350                & 4.178     & 25.674     & 61.109*** & 154.899** & 34.257  & -16.779  \\
             & (13.000)   & (14.752)  & (17.361)  & (16.052)   & (16.271)  & (19.058)  & (67.857) & (23.248)  & (26.090)  \\
First principal component   & 4.814***   & 7.745***   & 6.416***                 & 7.909***      & 7.640***      & 9.324***  & 18.978*** & 16.444***   & 10.058***   \\
             & (0.659)   & (0.527)  & (0.538)  & (0.644)   & (0.605)  & (0.834)  & (1.739)  & (0.976)  & (0.770)  \\
Constant     & -11.221 & -305.785***  & -57.771  & 40.313 & -81.538  & -273.069** & -808.902** & -131.549  & 174.291 \\
             & (74.681)  & (86.783) & (101.800) & (93.067)  & (95.084) & (111.175) & (404.593) & (137.064) & (154.931) \\ \hline
R2           & 0.11      & 0.16     & 0.18     & 0.22      & 0.21     & 0.16     & 0.11     & 0.25     & 0.22     \\
Observations & 447       & 927      & 736      & 554       & 662      & 772      & 874      & 934      & 594     \\ \hline
Mean school quality & 351.96 & 355.78 & 352.70 & 340.73 & 356.62 & 360.85 & 386.82 & 377.12 & 379.37 \\
             & (70.80)   & (61.36)  & (65.86)  & (63.18)   & (61.72)  & (66.15)  & (59.04)  & (74.86)  & (66.52)  \\
\hline
\end{tabular}
\end{table}

{\footnotesize \textit{Note}: Standard errors in parentheses. * p$<$0.10, ** p$<$0.05, *** p$<$0.01.
Mean school quality is the average total point score per pupil of the nearest secondary school,
with the within-region standard deviation in parentheses.}
\end{landscape}}
\end{center}

{\footnotesize \pagebreak }

\begin{center}
\begin{table}[h]
\caption{Quantile regression estimates for school quality demand at
different quartiles.}
\label{tab:q_iv}\centering%
\begin{tabular}{lccc}
\hline
& 25th percentile & 50th percentile & 75th percentile \\ \hline
$y-\theta _{1m(i)}$ & 0.00007*** & 0.00006*** & 0.00006*** \\ 
& (0.00001) & (0.00001) & (0.00001) \\ 
$\theta _{2m(i)}$ & -0.00017 & -0.00037*** & -0.00027*** \\ 
& (0.00011) & (0.00010) & (0.00010) \\ 
$\delta _{m(i)}$ & 0.00445*** & 0.00609*** & 0.00489*** \\ 
& (0.00081) & (0.00077) & (0.00077) \\ 
Constant & 5.69361*** & 5.80207*** & 5.91779*** \\ 
& (0.00805) & (0.00768) & (0.00760) \\ \hline
Observations & 6500 & 6500 & 6500 \\ \hline
\end{tabular}%
\end{table}

{\footnotesize \textit{Note}: Standard errors in parentheses. * p$<$0.10, **
p$<$0.05, *** p$<$0.01}

\newpage

\bigskip

\begin{table}[h]
\caption{Compensating variation estimates for quartiles of preference for
school quality, GBP.}
\label{tab:cv_nonp1}\centering%
\begin{tabular}{lccc}
\hline
& \multicolumn{3}{c}{Income percentile} \\ 
School quality percentile & 25th percentile & 50th percentile & 75th
percentile \\ \hline
25th percentile & -7.7681 & -7.4988 & -7.0970 \\ 
50th percentile & -3.7491 & -3.5227 & -3.1849 \\ 
75th percentile & 1.2724 & 1.4964 & 1.8305 \\ \hline
Observations & 6500 & 6500 & 6500 \\ \hline
\end{tabular}%
\end{table}
\end{center}

\newpage

\begin{center}
\textbf{References }
\end{center}

\begin{enumerate}
\item Andrews, J., Perera, N., Eyles, A., Sahlgren, G. H., Machin, S.,
Sandi, M., and Silva, O. (2017). The impact of academies on educational
outcomes. Education Policy Institute.

\item Bayer, P., Ferreira, F., \& McMillan, R. (2007). A unified framework
for measuring preferences for schools and neighborhoods. Journal of
Political Economy, 115(4), 588-638.

\item Bayer, P., McMillan, R., Murphy, A. and Timmins, C., 2016. A dynamic
model of demand for houses and neighborhoods. Econometrica, 84(3),
pp.893-942.

\item Black, S.E., 1999. Do better schools matter? Parental valuation of
elementary education. The Quarterly Journal of Economics, 114(2), pp.577-599.

\item Blomquist, S., Newey, W.K., Kumar, Anil. and Liang, C.Y., 2021. On
bunching and identification of the taxable income elasticity. Journal of
Political Economy, 129(8), pp.2320-2343.

\item Brown, J.N. and Rosen, H.S., 1982. On the estimation of structural
hedonic price models. Econometrica, Vol. 50, No. 3 (May, 1982), pp. 765-768.

\item Chernozhukov, V. and Hansen, C., 2005. An IV model of quantile
treatment effects. Econometrica, 73(1), pp.245-261.

\item Courant, R., John, F., Courant, R. and John, F., 1989. Multiple
Integrals. Introduction to Calculus and Analysis: Volume II, pp. 95-96.

\item Department for Communities and Local Government. (2018). English
Housing Survey, 2014-2016: Secure Access. [data collection]. 3rd Edition. UK
Data Service. SN: 8121, http://doi.org/10.5255/UKDA-SN-8121-3.

\item Department for Levelling Up, Housing and Communities (2022). English
Housing Survey Technical Report 2020-21. Section 5.95.

\item Diewert, W. E., 2003. Hedonic regressions. A consumer theory approach.
In Scanner data and price indexes (pp. 317-348). University of Chicago Press.

\item Duchini, E., Lavy, V., Machin, S. and Telhaj, S., 2023. School
Management Takeover, Leadership Change, and Personnel Policy (No. w31994).
National Bureau of Economic Research.

\item Ekeland, I., Heckman, J.J. and Nesheim, L., 2004. Identification and
estimation of hedonic models. Journal of Political Economy, 112(S1),
pp.S60-S109.

\item Eyles, A., Machin, S., \& Silva, O. (2017). Academies 2 -- The new
batch: The changing nature of academy schools in England. Fiscal Studies,
38(3), 437--454.

\item Gan, L., Ju, G. and Zhu, X., 2015. Nonparametric estimation of
structural labor supply and exact welfare change under nonconvex
piecewise-linear budget sets. Journal of Econometrics, 188(2), pp.526-544.

\item Hausman, J.A., 1981. Exact consumer's surplus and deadweight loss. The
American Economic Review, 71(4), pp.662-676.

\item Hausman, J.A. and Newey, W.K., 2016. Individual heterogeneity and
average welfare. Econometrica, 84(3), pp.1225-1248.

\item Heckman, J.J., Matzkin, R.L. and Nesheim, L., 2010. Nonparametric
Estimation of Nonadditive Hedonic Models, Econometrica, 78(5), pp.1569-1591.

\item Hoderlein, S. and Vanhems, A., 2018. Estimating the distribution of
welfare effects using quantiles. Journal of applied econometrics, 33(1),
pp.52-72.

\item Hussain, I. (2023). How school inspections affect house prices:
Evidence from England. Journal of Urban Economics, 136, 103506.

\item Kitamura, Y. and Stoye, J., 2018. Nonparametric analysis of random
utility models. Econometrica, 86(6), pp.1883-1909.

\item Lewbel, A., 2001. Demand Systems with and without Errors. American
Economic Review, 91(3), pp.611-618.

\item Machin, S. 2011. Houses and schools: Valuation of school quality
through the housing market. Labour Economics, 18(6), 723-729.

\item Machin, S. and Salvanes, K.G., 2016. Valuing school quality via a
school choice reform. The Scandinavian Journal of Economics, 118(1), pp.3-24.

\item MaCurdy, T., Green, D. and Paarsch, H., 1990. Assessing empirical
approaches for analyzing taxes and labor supply. Journal of Human Resources,
pp.415-490.

\item Mas-Colell, A., Whinston, M.D. and Jerry, R.G., 1995. Microeconomic
Theory. Oxford University Press.

\item McFadden, D.L., 2006. Revealed stochastic preference: a synthesis. In
Rationality and Equilibrium: A Symposium in Honor of Marcel K. Richter (pp.
1-20). Springer Berlin Heidelberg.

\item Moffitt, R., 1990. The econometrics of kinked budget constraints.
Journal of Economic Perspectives, 4(2), pp.119-139.

\item Pakes, A., 2003. A Reconsideration of Hedonic Price Indexes with an
Application to PC's. American Economic Review, 93(5), pp.1578-1596.

\item Rosen, S., 1974. Hedonic prices and implicit markets: product
differentiation in pure competition. Journal of Political Economy, 82(1),
pp.34-55.

\item Rosen, S. and Sanderson, A., 2001. Labour markets in professional
sports. The economic journal, 111(469), pp.47-68.

\item Saez, Emmanuel. 2010. "Do Taxpayers Bunch at Kink Points?" American
Economic Journal: Economic Policy 2 (3): 180--212.

\item Sheppard, S., 1999. Hedonic analysis of housing markets. Handbook of
regional and urban economics, 3, pp.1595-1635.

\item Spiegel, M.R., 2010. Schaum's Outline of Advanced Calculus.
McGraw-Hill.

\item Taylor, Angus, 1955. Advanced Calculus, Ginn and Company, New York
\end{enumerate}

\newpage

\appendix

\singlespacing

\section*{Appendix 1}

\begin{center}
\textbf{Proof of Lemma 1}
\end{center}

\begin{proof}
Let $Q_{\tau }\left( \mathbf{.,.}\right) $ be as in (\ref{1}), and $e\left( 
\mathbf{\theta },u\right) $ denote the expenditure function, i.e. the
solution to%
\begin{equation*}
Q_{\tau }\left( \mathbf{\theta },e\right) =u\text{.}
\end{equation*}%
This function is well defined since $Q_{\tau }\left( \mathbf{\theta }%
,e\right) $ is continuous and strictly increasing in $e$. Let $j=1$ and $k=2$
WLOG. Then, by definition%
\begin{equation*}
\begin{array}{l}
\frac{\partial }{\partial \theta _{1}}\left\{ Q_{\tau }\left( \mathbf{\theta 
},e\left( \mathbf{\theta },u\right) \right) \right\} =0 \\ 
\Longrightarrow \frac{\partial }{\partial \theta _{1}}Q_{\tau }\left( 
\mathbf{\theta },e\left( \mathbf{\theta },u\right) \right) +\frac{\partial }{%
\partial e}Q_{\tau }\left( \mathbf{\theta },e\left( \mathbf{\theta }%
,u\right) \right) \frac{\partial e\left( \mathbf{\theta },u\right) }{%
\partial \theta _{1}}=0 \\ 
\Longrightarrow \frac{\partial e\left( \mathbf{\theta },u\right) }{\partial
\theta _{1}}=-\frac{\frac{\partial }{\partial \theta _{1}}Q_{\tau }\left( 
\mathbf{\theta },e\left( \mathbf{\theta },u\right) \right) }{\frac{\partial 
}{\partial e}Q_{\tau }\left( \mathbf{\theta },e\left( \mathbf{\theta }%
,u\right) \right) }\overset{\text{by (\ref{9})}}{=}\left. \frac{\partial
P\left( s,\theta \right) }{\partial \theta _{1}}\right\vert _{s=q^{\tau
}\left( e\left( \mathbf{\theta },u\right) ,\theta \right) }%
\end{array}%
\end{equation*}%
Similarly,%
\begin{equation*}
\frac{\partial e\left( \mathbf{\theta },u\right) }{\partial \theta _{2}}%
=\left. \frac{\partial P\left( s,\theta \right) }{\partial \theta _{2}}%
\right\vert _{s=q^{\tau }\left( e\left( \mathbf{\theta },u\right) ,\theta
\right) }
\end{equation*}%
Thus we have that%
\begin{eqnarray}
\frac{\partial e\left( \mathbf{\theta },u\right) }{\partial \theta _{1}}
&=&\left. \frac{\partial P\left( s,\theta \right) }{\partial \theta _{1}}%
\right\vert _{s=q^{\tau }\left( e\left( \mathbf{\theta },u\right) ,\theta
\right) }  \label{S1} \\
\frac{\partial e\left( \mathbf{\theta },u\right) }{\partial \theta _{2}}
&=&\left. \frac{\partial P\left( s,\theta \right) }{\partial \theta _{2}}%
\right\vert _{s=q^{\tau }\left( e\left( \mathbf{\theta },u\right) ,\theta
\right) }  \label{S2}
\end{eqnarray}%
Now, differentiating (\ref{S1}) w.r.t. $\theta _{2}$, we get%
\begin{eqnarray}
\frac{\partial ^{2}e\left( \mathbf{\theta },u\right) }{\partial \theta
_{2}\partial \theta _{1}} &=&\frac{\partial }{\partial \theta _{2}}\left\{ 
\frac{\partial P\left( q^{\tau }\left( e\left( \mathbf{\theta },u\right)
,\theta \right) ,\theta \right) }{\partial \theta _{1}}\right\}  \notag \\
&=&\frac{\partial ^{2}P\left( q^{\tau }\left( e\left( \mathbf{\theta }%
,u\right) ,\theta \right) ,\theta \right) }{\partial \theta _{2}\partial q}%
\left\{ 
\begin{array}{l}
\frac{\partial q^{\tau }\left( e\left( \mathbf{\theta },u\right) ,\theta
\right) }{\partial y}\frac{\partial P\left( q^{\tau }\left( e\left( \mathbf{%
\theta },u\right) ,\theta \right) ,\theta \right) }{\partial \theta _{1}} \\ 
+\frac{\partial q^{\tau }\left( e\left( \mathbf{\theta },u\right) ,\theta
\right) }{\partial \theta _{1}}%
\end{array}%
\right\}  \notag \\
&&+\left\{ \frac{\partial ^{2}P\left( q^{\tau }\left( e\left( \mathbf{\theta 
},u\right) ,\theta \right) ,\theta \right) }{\partial \theta _{1}\partial
\theta _{2}}\right\}  \label{S3}
\end{eqnarray}%
Similarly, differentiating (\ref{S2}) w.r.t. $\theta _{1}$, we get%
\begin{eqnarray}
\frac{\partial ^{2}e\left( \mathbf{\theta },u\right) }{\partial \theta
_{1}\partial \theta _{2}} &=&\frac{\partial ^{2}P\left( q^{\tau }\left(
e\left( \mathbf{\theta },u\right) ,\theta \right) ,\theta \right) }{\partial
\theta _{1}\partial q}\left\{ 
\begin{array}{l}
\frac{\partial q^{\tau }\left( e\left( \mathbf{\theta },u\right) ,\theta
\right) }{\partial y}\frac{\partial P\left( q^{\tau }\left( e\left( \mathbf{%
\theta },u\right) ,\theta \right) ,\theta \right) }{\partial \theta _{2}} \\ 
+\frac{\partial q^{\tau }\left( e\left( \mathbf{\theta },u\right) ,\theta
\right) }{\partial \theta _{2}}%
\end{array}%
\right\}  \notag \\
&&+\left\{ \frac{\partial ^{2}P\left( q^{\tau }\left( e\left( \mathbf{\theta 
},u\right) ,\theta \right) ,\theta \right) }{\partial \theta _{1}\partial
\theta _{2}}\right\}  \label{S4}
\end{eqnarray}%
Equating (\ref{S3}) and (\ref{S4}), and evaluating at $y=e\left( \mathbf{%
\theta },u\right) $, we get the symmetry condition%
\begin{equation}
\begin{array}{c}
\frac{\partial ^{2}P\left( q^{\tau }\left( y,\theta \right) ,\theta \right) 
}{\partial \theta _{1}\partial q}\left\{ \frac{\partial q^{\tau }\left(
y,\theta \right) }{\partial y}\frac{\partial P\left( q^{\tau }\left(
y,\theta \right) ,\theta \right) }{\partial \theta _{2}}+\frac{\partial
q^{\tau }\left( y,\theta \right) }{\partial \theta _{2}}\right\} \\ 
=\frac{\partial ^{2}P\left( q^{\tau }\left( y,\theta \right) ,\theta \right) 
}{\partial \theta _{2}\partial q}\left\{ \frac{\partial q^{\tau }\left(
y,\theta \right) }{\partial y}\frac{\partial P\left( q^{\tau }\left(
y,\theta \right) ,\theta \right) }{\partial \theta _{1}}+\frac{\partial
q^{\tau }\left( y,\theta \right) }{\partial \theta _{1}}\right\}%
\end{array}
\label{S5}
\end{equation}
\end{proof}

\bigskip

\begin{center}
\textbf{Proof of Proposition 2}
\end{center}

\begin{proof}[Proof]
From (\ref{S1}), (\ref{p}), (\ref{d}), we have that%
\begin{equation}
\frac{\partial e\left( \mathbf{\theta },u\right) }{\partial \theta _{1}}%
\overset{\text{by (\ref{p})}}{=}1\Longrightarrow \frac{\partial ^{2}e\left(
\theta _{1},\theta _{2},u\right) }{\partial \theta _{2}\partial \theta _{1}}%
=0\text{.}  \label{zero}
\end{equation}%
On the other hand, by (\ref{S2})%
\begin{eqnarray}
&&%
\begin{array}{l}
\frac{\partial e\left( \mathbf{\theta },u\right) }{\partial \theta _{2}}%
\overset{\text{by (\ref{p})}}{=}\ln q^{\tau }\left( \theta _{1},\theta
_{2},e\left( \theta _{1},\theta _{2},u\right) \right) \\ 
\overset{\text{by (\ref{d})}}{=}r_{0}+r_{1}\times e\left( \theta _{1},\theta
_{2},u\right) +r_{2}\theta _{1}+r_{3}\theta _{2}%
\end{array}
\notag \\
&\Longrightarrow &\frac{\partial ^{2}e\left( \theta _{1},\theta
_{2},u\right) }{\partial \theta _{1}\partial \theta _{2}}=r_{2}+r_{1}\frac{%
\partial e\left( \theta _{1},\theta _{2},u\right) }{\partial \theta _{1}}%
\overset{\text{by (\ref{p})}}{=}r_{2}+r_{1}\text{.}  \label{z}
\end{eqnarray}

Now (\ref{zero}) and (\ref{z}) imply that $r_{1}+r_{2}=0$.
\end{proof}

\begin{center}
\bigskip

\textbf{Proof of Proposition 3}
\end{center}

\begin{proof}
Consider a price path $\theta \left( t\right) $, with $t\in \left[ 0,1\right]
$, such that $\theta _{1}\left( 0\right) =a_{1}$, $\theta _{2}\left(
0\right) =a_{2}$, $\theta _{1}\left( 1\right) =b_{1}$, $\theta _{2}\left(
1\right) =b_{2}$. By definition, the compensating variation at a generic
value of $t\in \left[ 0,1\right] $ is given by $T\left( t,y,\tau \right)
=e\left( \theta \left( t\right) ,\bar{u}\right) -y$, will satisfy%
\begin{equation}
Q_{\tau }\left( \theta \left( t\right) ,y+T\left( t,y,\tau \right) \right) =%
\bar{u}\text{ for all }t\text{,}  \label{2}
\end{equation}%
where the initial utility level $\bar{u}=Q_{\tau }\left(
a_{1},a_{2},y\right) $. Differentiating (\ref{2}) w.r.t. $t$ we have that
for all $t$,%
\begin{equation*}
\frac{d}{dt}Q_{\tau }\left( \theta \left( t\right) ,y+T\left( t,y,\tau
\right) \right) =0\text{, i.e.}
\end{equation*}%
for all $t$,%
\begin{equation*}
\displaystyle\sum\limits_{j}\frac{\partial Q_{\tau }\left( \theta \left( t\right)
,y+T\left( t,y,\tau \right) \right) }{\partial \theta _{j}}\frac{d\theta
_{j}\left( t\right) }{dt}+\frac{\partial Q_{\tau }\left( \theta \left(
t\right) ,y+T\left( t,y,\tau \right) \right) }{\partial y}\frac{dT\left(
t,y,\tau \right) }{dt}=0\text{.}
\end{equation*}%
This implies%
\begin{eqnarray}
\frac{dT\left( t,y,\tau \right) }{dt} &=&-\displaystyle\sum\limits_{j}\frac{\frac{%
\partial Q_{\tau }\left( \theta \left( t\right) ,y+T\left( t,y,\tau \right)
\right) }{\partial \theta _{j}}}{\frac{\partial Q_{\tau }\left( \theta
\left( t\right) ,y+T\left( t,y,\tau \right) \right) }{\partial y}}\frac{%
d\theta _{j}\left( t\right) }{dt}  \notag \\
&&\overset{\text{by (\ref{6})}}{=}\displaystyle\sum\limits_{j}\frac{d\theta _{j}\left(
t\right) }{dt}\times \left. \frac{\partial P\left( S,\theta \right) }{%
\partial \theta _{j}}\right\vert _{\theta =\theta \left( t\right) ,s=q^{\tau
}\left( \theta \left( t\right) ,e\left( \theta \left( t\right) ,\bar{u}%
\right) \right) }  \label{ode}
\end{eqnarray}%
The second term on the RHS of (\ref{ode}) is identifiable from the data
across many markets, since $e\left( \theta \left( t\right) ,\bar{u}\right)
=y+T\left( t,y,\tau \right) $. So finding the compensating variation for a
change in $\theta $ from $a$ to $b$ for an individual at income $y$ and
whose $\eta $ equals its $\tau $th quantile is equivalent to finding $%
T\left( 1,y,\tau \right) $, where $T\left( t,y,\tau \right) $ solves (\ref%
{ode}) with the initial condition $T\left( 0,y,\tau \right) =0$. Observe that%
\begin{eqnarray}
T\left( 1.y,\tau \right) -T\left( 0,y,\tau \right) &=&\int_{0}^{1}\frac{%
dT\left( t,y,\tau \right) }{dt}dt  \notag \\
&=&\int_{0}^{1}\left\{ \displaystyle\sum\limits_{j}\frac{d\theta _{j}\left( t\right) }{dt%
}\times \frac{\partial P\left( \theta \left( t\right) ,q^{\tau }\left(
\theta \left( t\right) ,e\left( \theta \left( t\right) ,\bar{u}\right)
\right) \right) }{\partial \theta _{j}}\right\} dt  \notag \\
&=&\int_{\mathcal{G}}\displaystyle\sum\limits_{j}\frac{\partial P\left( \theta ,q^{\tau
}\left( \theta ,e\left( \theta ,\bar{u}\right) \right) \right) }{\partial
\theta _{j}}d\theta _{j}\text{.}  \label{l}
\end{eqnarray}%
The final expression on the RHS is a line integral of the vector field $%
\left( \frac{\partial P\left( \theta ,q^{\tau }\left( \theta ,e\left( \theta
,\bar{u}\right) \right) \right) }{\partial \theta _{1}}\text{, }\frac{%
\partial P\left( \theta ,q^{\tau }\left( \theta ,e\left( \theta ,\bar{u}%
\right) \right) \right) }{\partial \theta _{2}}\right) $ along the path $%
\mathcal{G=}\left\{ \theta _{1}\left( t\right) ,\theta _{2}\left( t\right)
\right\} $, $t\in \left[ 0,1\right] $ connecting the points $\left(
a_{1},a_{2}\right) $ and $\left( b_{1},b_{2}\right) $. The symmetry
condition (\ref{S5}) and the gradient theorem for line integrals (cf.
Spiegel 2010, Sec 10.6; Courant and John 1989, Sec 1.10) then imply that the
value of (\ref{l}) is path-independent, i.e. its value does not depend on
the path $\mathcal{G}$. Thus the compensating variation $T\left( 1.y\right) $
is well-defined.

In particular, given the specification (\ref{p}) and (\ref{d}), equation (%
\ref{ode}) reduces to the ordinary differential equation%
\begin{eqnarray*}
&&\frac{dT\left( t,y,\tau \right) }{dt} \\
&&\overset{\text{(by \ref{d})}}{=}\frac{d\theta _{1}\left( t\right) }{dt}+%
\frac{d\theta _{2}\left( t\right) }{dt}\times \left( r_{0}+r_{1}\left(
y+T\left( t,y,\tau \right) \right) +r_{2}\theta _{1}\left( t\right)
+r_{3}\theta _{2}\left( t\right) \right) \\
&&\overset{\text{by }r_{2}+r_{1}=0}{=}\frac{d\theta _{1}\left( t\right) }{dt}%
+\frac{d\theta _{2}\left( t\right) }{dt}\times \left( r_{0}+r_{1}\left(
y+T\left( t,y,\tau \right) -\theta _{1}\left( t\right) \right) +r_{3}\theta
_{2}\left( t\right) \right) \\
&\Leftrightarrow &\frac{dT\left( t,y,\tau \right) }{dt}-r_{1}\frac{d\theta
_{2}\left( t\right) }{dt}\times T\left( t,y,\tau \right) \\
&=&\frac{d\theta _{1}\left( t\right) }{dt}+\frac{d\theta _{2}\left( t\right) 
}{dt}\times \left( r_{0}+r_{1}\left( y-\theta _{1}\left( t\right) \right)
\right) +r_{3}\frac{d\theta _{2}\left( t\right) }{dt}\times \theta
_{2}\left( t\right)
\end{eqnarray*}%
This implies%
\begin{eqnarray*}
&&\frac{dT\left( t,y,\tau \right) }{dt}-r_{1}\frac{d\theta _{2}\left(
t\right) }{dt}\times T\left( t,y,\tau \right) \\
&=&\frac{d\theta _{1}\left( t\right) }{dt}+\frac{d\theta _{2}\left( t\right) 
}{dt}\times \left( r_{0}+r_{1}\left( y-\theta _{1}\left( t\right) \right)
\right) +r_{3}\frac{d\theta _{2}\left( t\right) }{dt}\times \theta
_{2}\left( t\right)
\end{eqnarray*}%
This linear ODE can be solved using the method of integrating factors as%
\begin{eqnarray*}
&&\frac{dT\left( t,y,\tau \right) }{dt}e^{-r_{1}\theta _{2}\left( t\right)
}-r_{1}\frac{d\theta _{2}\left( t\right) }{dt}e^{-r_{1}\theta _{2}\left(
t\right) }\times T\left( t,y,\tau \right) \\
&=&\frac{d\theta _{1}\left( t\right) }{dt}e^{-r_{1}\theta _{2}\left(
t\right) }+e^{-r_{1}\theta _{2}\left( t\right) }\frac{d\theta _{2}\left(
t\right) }{dt}\times \left( r_{0}+r_{1}\left( y-\theta _{1}\left( t\right)
\right) \right) \\
&&+r_{3}\frac{d\theta _{2}\left( t\right) }{dt}\times \theta _{2}\left(
t\right) e^{-r_{1}\theta _{2}\left( t\right) } \\
&=&\frac{d\theta _{1}\left( t\right) }{dt}e^{-r_{1}\theta _{2}\left(
t\right) }-r_{1}e^{-r_{1}\theta _{2}\left( t\right) }\frac{d\theta
_{2}\left( t\right) }{dt}\times \theta _{1}\left( t\right) \\
&&+\left( r_{0}+r_{1}y\right) e^{-r_{1}\theta _{2}\left( t\right) }\frac{%
d\theta _{2}\left( t\right) }{dt}+r_{3}\frac{d\theta _{2}\left( t\right) }{dt%
}\times \theta _{2}\left( t\right) e^{-r_{1}\theta _{2}\left( t\right) }
\end{eqnarray*}%
This implies%
\begin{eqnarray*}
\frac{d}{dt}\left\{ T\left( t,y,\tau \right) e^{-r_{1}\theta _{2}\left(
t\right) }\right\} &=&\frac{d}{dt}\left\{ \theta _{1}\left( t\right)
e^{-r_{1}\theta _{2}\left( t\right) }\right\} \\
&&+e^{-r_{1}\theta _{2}\left( t\right) }\frac{d\theta _{2}\left( t\right) }{%
dt}\left( r_{0}+r_{1}y\right) \\
&&+r_{3}\frac{d\theta _{2}\left( t\right) }{dt}\times \theta _{2}\left(
t\right) e^{-r_{1}\theta _{2}\left( t\right) }
\end{eqnarray*}%
Integrating both sides, we get that%
\begin{eqnarray}
T\left( t,y,\tau \right) e^{-r_{1}\theta _{2}\left( t\right) }
&=&Tons+\theta _{1}\left( t\right) e^{-r_{1}\theta _{2}\left( t\right)
}-\left( r_{0}+r_{1}y\right) \frac{e^{-r_{1}\theta _{2}\left( t\right) }}{%
r_{1}}  \notag \\
&&-r_{3}\theta _{2}\left( t\right) \frac{e^{-r_{1}\theta _{2}\left( t\right)
}}{r_{1}}-\frac{e^{-r_{1}\theta _{2}\left( t\right) }}{r_{1}^{2}}r_{3}\text{,%
}  \label{X}
\end{eqnarray}%
where \textquotedblleft $Tons$\textquotedblright\ denotes a constant. This
implies%
\begin{equation*}
T\left( t,y,\tau \right) =Tons\times e^{r_{1}\theta _{2}\left( t\right)
}+\theta _{1}\left( t\right) -\left( r_{0}+r_{1}y\right) \frac{1}{r_{1}}%
-r_{3}\theta _{2}\left( t\right) \frac{1}{r_{1}}-\frac{r_{3}}{r_{1}^{2}}
\end{equation*}%
Applying the boundary condition $T\left( 0,y,\tau \right) =0$, and $\theta
_{1}\left( 0\right) =a_{1}$, $\theta _{2}\left( 0\right) =a_{2}$, we get%
\begin{equation*}
Cons=e^{-r_{1}a_{2}}\left\{ -a_{1}+\left( r_{0}+r_{1}y\right) \frac{1}{r_{1}}%
+a_{2}\frac{r_{3}}{r_{1}}+\frac{r_{3}}{r_{1}^{2}}\right\} \text{.}
\end{equation*}%
Replacing in (\ref{X}), and evaluating at $t=1$, using $\theta _{1}\left(
1\right) =b_{1}$, $\theta _{2}\left( 1\right) =b_{2}$, we get%
\begin{eqnarray}
T\left( 1,y,\tau \right) &=&e^{r_{1}\left( b_{2}-a_{2}\right) }\left\{
-a_{1}+\frac{r_{0}}{r_{1}}+y+\frac{a_{2}r_{3}}{r_{1}}+\frac{r_{3}}{r_{1}^{2}}%
\right\}  \notag \\
&&+b_{1}-\frac{r_{0}}{r_{1}}-y-\frac{r_{3}b_{2}}{r_{1}}-\frac{r_{3}}{%
r_{1}^{2}}  \label{CV}
\end{eqnarray}%
It is clear that $T\left( 1,y,\tau \right) $ does not depend on the path
from $\left( a_{1},a_{2}\right) $ to $\left( b_{1},b_{2}\right) $ because
the exact form of $\theta _{1}\left( t\right) ,\theta _{2}\left( t\right) $
as functions of $t$ were never used to derive (\ref{CV}).

\newpage
\end{proof}

\begin{longtable}{lrr}
\caption{Descriptive statistics for matched property-school observations.} \label{tab:desc} \\
\hline
\multicolumn{1}{c}{Variable}  & \multicolumn{1}{c}{Mean} & \multicolumn{1}{c}{SD} \\
\hline
\endhead
\hline
\endfoot
Total weekly rent                                   & 119.69 & 61.37  \\
Net household income including savings and benefits & 415.72 & 220.47 \\
Total average point score per pupil                 & 363.98 & 67.06  \\
Logarithm of total average point score per pupil    & 5.88   & 0.20   \\
Floor area, sqm                                     & 66.72  & 23.28  \\
Number of floors:  && \\
\quad 1                                 & 0.10   & 0.30   \\
\quad 2                                 & 0.67   & 0.47   \\
\quad 3                                 & 0.13   & 0.34   \\
\quad 4                                 & 0.05   & 0.22   \\
\quad 5 or more                         & 0.05   & 0.22   \\
Dwelling type: && \\
\quad end terrace                          & 0.12   & 0.32   \\
\quad mid terrace                          & 0.22   & 0.41   \\
\quad  semi detached                        & 0.20   & 0.40   \\
\quad  detached                             & 0.02   & 0.14   \\
\quad  bungalow                             & 0.10   & 0.29   \\
\quad converted flat                       & 0.05   & 0.21   \\
\quad  purpose built flat, low rise         & 0.28   & 0.45   \\
\quad  purpose built flat, high rise        & 0.03   & 0.17   \\
Dwelling age: && \\
\quad pre 1850                              & 0.01   & 0.11   \\
\quad 1850 to 1899                          & 0.06   & 0.24   \\
\quad 1900 to 1918                          & 0.06   & 0.24   \\
\quad 1919 to 1944                          & 0.13   & 0.34   \\
\quad 1945 to 1964                          & 0.26   & 0.44   \\
\quad 1965 to 1974                          & 0.16   & 0.37   \\
\quad 1975 to 1980                          & 0.08   & 0.26   \\
\quad 1981 to 1990                          & 0.08   & 0.28   \\
\quad 1991 to 2002                          & 0.08   & 0.27   \\
\quad post 2002                             & 0.07   & 0.26   \\
Number of bedrooms: && \\
\quad 1                               & 0.23   & 0.42   \\
\quad 2                               & 0.36   & 0.48   \\
\quad 3                               & 0.35   & 0.48   \\
\quad 4                               & 0.04   & 0.20   \\
\quad 5 or more                       & 0.01   & 0.09   \\
Number of living rooms: && \\
\quad 1                           & 0.01   & 0.09   \\
\quad 2                           & 0.88   & 0.33   \\
\quad 3                           & 0.11   & 0.31   \\
\quad 4                           & 0.01   & 0.07   \\
\quad 5 or more                   & 0.00   & 0.04   \\
Number of bathrooms: && \\
\quad 1                              & 0.94   & 0.23   \\
\quad 2                              & 0.05   & 0.22   \\
\quad 3 or more                      & 0.01   & 0.08   \\
Tenure type: && \\
\quad Private rented                         & 0.32   & 0.47   \\
\quad Local Authority                        & 0.29   & 0.45   \\
\quad Housing Association                    & 0.39   & 0.49   \\
Housing benefits: && \\
\quad Yes                               & 0.54   & 0.50   \\
\quad No                                & 0.46   & 0.50   \\
Deprivation decile: && \\
\quad 1 - Most deprived               & 0.23   & 0.42   \\
\quad 2                               & 0.17   & 0.38   \\
\quad 3                               & 0.14   & 0.34   \\
\quad 4                               & 0.11   & 0.32   \\
\quad 5                               & 0.09   & 0.28   \\
\quad 6                               & 0.08   & 0.27   \\
\quad 7                               & 0.07   & 0.25   \\
\quad 8                               & 0.05   & 0.22   \\
\quad 9                               & 0.04   & 0.19   \\
\quad 10 - Least deprived             & 0.03   & 0.16   \\
Region: && \\
\quad North East                                  & 0.07   & 0.25   \\
\quad North West                                  & 0.14   & 0.35   \\
\quad Yorkshire and the Humber                    & 0.11   & 0.32   \\
\quad East Midlands                               & 0.09   & 0.28   \\
\quad West Midlands                               & 0.10   & 0.30   \\
\quad East                                        & 0.12   & 0.32   \\
\quad London                                      & 0.13   & 0.34   \\
\quad South East                                  & 0.14   & 0.35   \\
\quad South West                                  & 0.09   & 0.29   \\
\hline
Observations                                        & 6,500   &        \\ 
\end{longtable}

\newpage

\begin{table}[h]
\caption{Number of support points of school quality, by market.}
\label{tab:supp}\centering%
\begin{tabular}{lccc}
\hline
Market & No. of Support Points & No. of Distinct & Observations \\ 
& of School Quality & Schools & (No. of Rented Properties) \\ \hline
North East & 118 & 120 & 447 \\ 
North West & 305 & 337 & 927 \\ 
Yorkshire and the Humber & 223 & 236 & 736 \\ 
East Midlands & 200 & 211 & 554 \\ 
West Midlands & 270 & 287 & 662 \\ 
East & 253 & 268 & 772 \\ 
London & 291 & 308 & 874 \\ 
South East & 340 & 371 & 934 \\ 
South West & 217 & 227 & 594 \\ \hline
\end{tabular}%
\end{table}

\newpage

\begin{longtable}{lr}
\caption{Correlation between floor area and other property characteristics.} \label{tab:pc_cor} \\
\hline
Variable & Floor area, sqm\\
\hline
\endhead

\hline
\endfoot
Floor area squared                           & 0.891        \\
Number of floors: & \\
\quad 2                          & 0.168        \\
\quad 3                          & 0.009        \\
\quad 4                          & -0.029      \\
\quad 5 or more                  & -0.076      \\
Dwelling type: & \\
\quad mid terrace                   & 0.177        \\
\quad semi detached                 & 0.213        \\
\quad detached                      & 0.308        \\
\quad bungalow                      & -0.195       \\
\quad converted flat                & -0.089       \\
\quad purpose built flat, low rise  & -0.347       \\
\quad purpose built flat, high rise & -0.049      \\
Dwelling age: & \\
\quad 1850 to 1899                   & 0.066       \\
\quad 1900 to 1918                   & 0.060       \\
\quad 1919 to 1944                   & 0.044       \\
\quad 1945 to 1964                   & 0.014       \\
\quad 1965 to 1974                   & -0.052      \\
\quad 1975 to 1980                   & -0.072      \\
\quad 1981 to 1990                   & -0.116       \\
\quad 1991 to 2002                   & -0.006      \\
\quad post 2002                      & 0.052       \\
Number of bedrooms: & \\
\quad 2                        & -0.134       \\
\quad 3                        & 0.367        \\
\quad 4                        & 0.361        \\
\quad 5 or more                & 0.337        \\
Number of living rooms: & \\
\quad 2                    & -0.291       \\
\quad 3                    & 0.292        \\
\quad 4                    & 0.156        \\
\quad 5 or more            & 0.037       \\
Number of bathrooms: & \\
\quad 2                       & 0.305        \\
\quad 3 or more               & 0.244        \\
Deprivation decile: & \\
\quad 2                        & -0.033      \\
\quad 3                        & -0.021     \\
\quad 4                        & -0.016     \\
\quad 5                        & 0.025      \\
\quad 6                        & 0.032       \\
\quad 7                        & -0.014      \\
\quad 8                        & -0.009      \\
\quad 9                        & 0.039       \\
\quad 10 - Least deprived      & 0.060      \\
Tenure type: & \\
\quad Local Authority                 & -0.074      \\
\quad Housing Association             & -0.079      \\
Housing benefits: & \\
\quad No                         & 0.127       \\
\hline
Observations  & 6,500\\ 
\end{longtable}

\newpage

\begin{longtable}{llc}
\caption{Principal component loadings and unexplained variance for property characteristics.} \label{tab:pc_loadings} \\
\hline
\textbf{Variable} & \textbf{Loadings on PC1} & \textbf{Unexplained Variance} \\
\hline
\endhead
\hline
\endfoot
Floor area, sqm                              & 0.407  & 0.356  \\
Floor area - squared, sqm                    & 0.357  & 0.506  \\
Number of floors: && \\
\quad 2                          & 0.238  & 0.780  \\
\quad 3                          & -0.085 & 0.972  \\
\quad 4                          & -0.086 & 0.971  \\
\quad 5 or more                  & -0.120 & 0.944  \\
Dwelling type: && \\
\quad mid terrace                   & 0.131  & 0.933  \\
\quad semi detached                 & 0.196  & 0.851  \\
\quad detached                      & 0.194  & 0.853  \\
\quad bungalow                      & -0.127 & 0.937  \\
\quad converted flat                & -0.037 & 0.995  \\
\quad purpose built flat, low rise  & -0.263 & 0.732  \\
\quad purpose built flat, high rise & -0.093 & 0.966  \\
Dwelling age: && \\
\quad 1850 to 1899                   & 0.049  & 0.991  \\
\quad 1900 to 1918                   & 0.066  & 0.983  \\
\quad 1919 to 1944                   & 0.091  & 0.968  \\
\quad 1945 to 1964                   & 0.010  & 1.000  \\
\quad 1965 to 1974                   & -0.083 & 0.973  \\
\quad 1975 to 1980                   & -0.064 & 0.984  \\
\quad 1981 to 1990                   & -0.063 & 0.985  \\
\quad 1991 to 2002                   & 0.005  & 1.000  \\
\quad post 2002                      & -0.013 & 0.999  \\
Number of bedrooms: && \\
\quad 2                        & -0.167  & 0.892  \\
\quad 3                        & 0.267  & 0.723  \\
\quad 4                        & 0.172  & 0.886  \\
\quad 5 or more                & 0.153  & 0.909  \\
Number of living rooms: && \\ 
\quad 2                    & -0.296  & 0.659  \\
\quad 3                    & 0.289  & 0.676  \\
\quad 4                    & 0.110  & 0.953  \\
\quad 5 or more            & 0.027  & 0.997  \\
Number of bathrooms: && \\
\quad 2                       & 0.165  & 0.895  \\
\quad 3 or more               & 0.118  & 0.946  \\
Deprivation decile: && \\
\quad 2                        & -0.047 & 0.991  \\
\quad 3                        & -0.021 & 0.998  \\
\quad 4                        & 0.002  & 1.000  \\
\quad 5                        & 0.013  & 0.999  \\
\quad 6                        & 0.028  & 0.997  \\
\quad 7                        & 0.008  & 1.000  \\
\quad 8                        & 0.011  & 1.000  \\
\quad 9                        & 0.030  & 0.997  \\
\quad 10 - Least deprived      & 0.056  & 0.988  \\
Tenure type: && \\
\quad Local Authority                 & -0.072  & 0.980  \\
\quad Housing Association             & -0.056  & 0.988  \\
Housing benefits: && \\
\quad No                         & 0.094  & 0.966  \\
\hline
Observations & 6,500 & \\ 
\end{longtable}

\newpage

\begin{center}
\begin{table}[h]
\caption{{}Quantile instrumental variable regression estimates for school
quality demand at different quartiles. Estimates with IV\ for income.}
\label{tab:q_2iv}\centering%
\begin{tabular}{lccc}
\hline
& 25th percentile & 50th percentile & 75th percentile \\ \hline
$y-\theta _{1m(i)}$ & 0.00015*** & 0.00016*** & 0.00010** \\ 
& (0.00004) & (0.00004) & (0.00005) \\ 
$\theta _{2m(i)}$ & -0.00066*** & -0.00098*** & -0.00059** \\ 
& (0.00023) & (0.00024) & (0.00028) \\ 
$\delta _{m(i)}$ & 0.00403*** & 0.00532*** & 0.00481*** \\ 
& (0.00086) & (0.00087) & (0.00087) \\ 
Constant & 5.66782*** & 5.77539*** & 5.90530*** \\ 
& (0.01334) & (0.01206) & (0.01304) \\ \hline
Observations & 6,500 & 6,500 & 6,500 \\ \hline
\end{tabular}%
\end{table}

{\footnotesize \textit{Note}: To address measurement error in reported
income, the term }${\footnotesize y-}\theta _{{\footnotesize 1m(i)}}$%
{\footnotesize \ in the model is instrumented with }${\footnotesize s-}%
\theta _{{\footnotesize 1m(i)}}${\footnotesize , where }${\footnotesize s}$%
{\footnotesize \ is the level of savings reported by the household.
F-statistics for the first stage from a linear regression model for the mean
equals 552.678. Standard errors in parentheses. * p$<$0.10, ** p$<$0.05, ***
p$<$0.01}

\newpage

\bigskip

\begin{table}[h]
\caption{Compensating variation estimates for quartiles of preference for
school quality, GBP. Estimates with IV for income.}
\label{tab:cv_nonp1_2iv}\centering%
\begin{tabular}{lccc}
\hline
& \multicolumn{3}{c}{Income percentile} \\ 
School quality percentile & 25th percentile & 50th percentile & 75th
percentile \\ \hline
25th percentile & -8.178 & -7.572 & -6.667 \\ 
50th percentile & -3.958 & -3.296 & -2.307 \\ 
75th percentile & 1.118 & 1.543 & 2.177 \\ \hline
\end{tabular}%
\end{table}

\begin{table}[h]
\caption{Quantile regression estimates for school quality demand at
different quartiles, by subsample.}
\label{tab:q_iv_subsamples}\centering%
\begin{tabular}{lccc}
\hline
& 25th percentile & 50th percentile & 75th percentile \\ \hline
\multicolumn{4}{l}{\textit{Panel A. All renters}} \\ 
$y-\theta _{1m(i)}$ & 0.00007*** & 0.00006*** & 0.00006*** \\ 
& (0.00001) & (0.00001) & (0.00001) \\ 
$\theta _{2m(i)}$ & -0.00017 & -0.00037*** & -0.00027*** \\ 
& (0.00011) & (0.00010) & (0.00010) \\ 
$\delta _{m(i)}$ & 0.00445*** & 0.00609*** & 0.00489*** \\ 
& (0.00081) & (0.00077) & (0.00077) \\ 
Constant & 5.69361*** & 5.80207*** & 5.91779*** \\ 
& (0.00805) & (0.00768) & (0.00760) \\ 
Observations & 6500 & 6500 & 6500 \\ \hline
\multicolumn{4}{l}{\textit{Panel B. Private renters}} \\ 
$y-\theta _{1m(i)}$ & 0.00009*** & 0.00006*** & 0.00007*** \\ 
& (0.00002) & (0.00002) & (0.00002) \\ 
$\theta _{2m(i)}$ & -0.00043** & -0.00053*** & -0.00032** \\ 
& (0.00017) & (0.00016) & (0.00015) \\ 
$\delta _{m(i)}$ & 0.00325** & 0.00659*** & 0.00309** \\ 
& (0.00142) & (0.00136) & (0.00129) \\ 
Constant & 5.71791*** & 5.81660*** & 5.94614*** \\ 
& (0.01452) & (0.01390) & (0.01318) \\ 
Observations & 2063 & 2063 & 2063 \\ \hline
\multicolumn{4}{l}{\textit{Panel C. Private renters with children}} \\ 
$y-\theta _{1m(i)}$ & 0.00011*** & 0.00007** & 0.00005** \\ 
& (0.00003) & (0.00003) & (0.00003) \\ 
$\theta _{2m(i)}$ & -0.00043* & -0.00053** & -0.00033 \\ 
& (0.00025) & (0.00023) & (0.00024) \\ 
$\delta _{m(i)}$ & 0.00289 & 0.00748*** & 0.00579*** \\ 
& (0.00203) & (0.00184) & (0.00192) \\ 
Constant & 5.70883*** & 5.80263*** & 5.92338*** \\ 
& (0.02144) & (0.01938) & (0.02030) \\ 
Observations & 947 & 947 & 947 \\ \hline
\end{tabular}%
\par
\medskip {\footnotesize \textit{Note}: Panel A repeats the estimates of
Table \ref{tab:q_iv}. All subsamples are imposed at Step 3 of Section \ref%
{Calculation_nonp}; the hedonic frontiers of Step 2 are estimated on the
full regional samples throughout. Standard errors in parentheses. * p$<$%
0.10, ** p$<$0.05, *** p$<$0.01}
\end{table}

\begin{table}[h]
\caption{Compensating variation estimates for quartiles of preference for
school quality by subsample, GBP.}
\label{tab:cv_subsamples}\centering%
\begin{tabular}{lccc}
\hline
& \multicolumn{3}{c}{Income percentile} \\ 
School quality percentile & 25th percentile & 50th percentile & 75th
percentile \\ \hline
\multicolumn{4}{l}{\textit{Panel A. All renters}} \\ 
25th percentile & -7.7681 & -7.4988 & -7.0970 \\ 
50th percentile & -3.7491 & -3.5227 & -3.1849 \\ 
75th percentile & 1.2724 & 1.4964 & 1.8305 \\ 
Observations & 6500 & 6500 & 6500 \\ \hline
\multicolumn{4}{l}{\textit{Panel B. Private renters}} \\ 
25th percentile & -6.6781 & -6.2946 & -5.7226 \\ 
50th percentile & -3.3089 & -3.0556 & -2.6778 \\ 
75th percentile & 2.7089 & 3.0019 & 3.4389 \\ 
Observations & 2063 & 2063 & 2063 \\ \hline
\multicolumn{4}{l}{\textit{Panel C. Private renters with children}} \\ 
25th percentile & -6.8251 & -6.3896 & -5.7400 \\ 
50th percentile & -3.8160 & -3.5442 & -3.1387 \\ 
75th percentile & 1.3834 & 1.6038 & 1.9325 \\ 
Observations & 947 & 947 & 947 \\ \hline
\end{tabular}%
\par
\medskip {\footnotesize \textit{Note}: Panel A repeats the estimates of
Table \ref{tab:cv_nonp1}. Each estimate is computed over the full subsample
at the stated quantile of income.}
\end{table}
\end{center}

\newpage

\section{Appendix 2: Nonparametric Roy's Identity and Slutsky Symmetry}

In this section, we treat the hedonic price schedule itself as the object
characterizing the budget frontier. Thus, rather than assuming that the
equilibrium price function is indexed by a finite-dimensional parameter
vector, let $P(s)=\theta (s)$, where $\theta $ belongs to a class $\Theta $
of sufficiently smooth functions defined on the support ${\mathcal{S}}$ of
the amenity $S$. For concreteness, we may take $\Theta $ to be an open
subset of $C^{2}({\mathcal{S}})$, equipped with the usual $C^{2}$ norm.
Different markets may therefore be characterized by different price
schedules $\theta \in \Theta $.

For a functional $F:\Theta \rightarrow {\mathbb{R}}$, let 
\begin{equation*}
D_{\theta }F(\theta )[h]=\left .\frac{d}{d\varepsilon }F(\theta +\varepsilon
h)\right |_{\varepsilon =0}
\end{equation*}
denote its directional derivative at $\theta $ in direction $h$, whenever
this derivative exists. The direction $h$ represents an arbitrary smooth
perturbation of the hedonic price schedule.

\subsection{Functional Roy's Identity and Slutsky Symmetry}

For simplicity of exposition, we restrict attention to the case where $S$ is
the only amenity of interest. An individual with income $y$, preference
heterogeneity $\eta $, and hedonic price schedule $\theta $ chooses $s$ to
maximize $U(s,y-\theta (s);\eta ).$

Denote the resulting structural demand by $S^{\ast }(y,\theta ,\eta )$. Then
at an interior optimum, 
\begin{equation}
U_{s}(S^{\ast },y-\theta (S^{\ast });\eta )-U_{c}(S^{\ast },y-\theta
(S^{\ast });\eta )\theta ^{\prime }\left( S^{\ast }\right) )=0.
\end{equation}

Define indirect utility by 
\begin{equation}
V(y,\theta ,\eta )=U\left (S^*(y,\theta ,\eta ),y-\theta (S^*(y,\theta ,\eta
));\eta \right ).
\end{equation}

By the envelope theorem, 
\begin{equation}
\frac{\partial }{\partial y}V(y,\theta ,\eta )=U_{c}\left( S^{\ast
}(y,\theta ,\eta ),y-\theta (S^{\ast }(y,\theta ,\eta ));\eta \right) >0.
\label{A4}
\end{equation}

Now consider an arbitrary perturbation $h\in C^{2}({\mathcal{S}})$ of the
price schedule. The functional envelope theorem gives 
\begin{equation}
D_{\theta }V(y,\theta ,\eta )[h]=-U_{c}\left( S^{\ast },y-\theta (S^{\ast
});\eta \right) h(S^{\ast }).  \label{A5}
\end{equation}

Combining (\ref{A4}) and (\ref{A5}) yields the functional analogue of Roy's
identity:

\begin{equation}
D_{\theta }V(y,\theta ,\eta )[h]+V_{y}(y,\theta ,\eta )\,h(S^{\ast
}(y,\theta ,\eta ))=0  \label{A6}
\end{equation}%
for every admissible direction $h$.

Notice that the finite-dimensional formulation is nested within (\ref{A6}).
If 
\begin{equation*}
\theta (s)=P(s;\beta ),\qquad \beta =(\beta _{1},\ldots ,\beta _{J}),
\end{equation*}%
and we choose the perturbation $h_{j}(s)=\frac{\partial P(s;\beta )}{%
\partial \beta _{j}}$, then (\ref{A6}) becomes 
\begin{equation*}
\frac{\partial V}{\partial \beta _{j}}+V_{y}\frac{\partial P(S^{\ast };\beta
)}{\partial \beta _{j}}=0,
\end{equation*}%
which is the finite-dimensional Roy identity derived previously.

\subsubsection*{Functional Slutsky symmetry}

For two arbitrary perturbations $h,k\in C^{2}({\mathcal{S}})$, define the
compensated directional response of structural demand in direction $h$ by 
\begin{equation*}
{\mathcal{S}}_{h}(y,\theta ,\eta )=D_{\theta }S^{\ast }(y,\theta ,\eta )[h]+%
\frac{\partial S^{\ast }(y,\theta ,\eta )}{\partial y}h\left( S^{\ast
}(y,\theta ,\eta )\right) .
\end{equation*}

We now derive the functional Slutsky symmetry condition. For any admissible
direction $h$, the functional Roy identity is 
\begin{equation*}
D_{\theta}V(y,\theta,\eta)[h] + V_y(y,\theta,\eta)h(S^{\ast}) =0.
\end{equation*}
Assume that $V$ is sufficiently smooth that its second Frechet derivative
with respect to $\theta$ exists and is symmetric, and that $S^{\ast}$ is
differentiable with respect to $y$ and $\theta$.

Differentiate the functional Roy identity in direction $h$ with respect to $%
\theta$ in another admissible direction $k$. Since $h$ is fixed but is
evaluated at the endogenous choice $S^{\ast}$, we obtain 
\begin{equation*}
D_{\theta\theta}^{2}V[h,k] + D_{\theta}V_y[k]h(S^{\ast}) + V_y
h^{\prime}(S^{\ast})D_{\theta}S^{\ast}[k] =0.
\end{equation*}
Interchanging $h$ and $k$ gives 
\begin{equation*}
D_{\theta\theta}^{2}V[k,h] + D_{\theta}V_y[h]k(S^{\ast}) + V_y
k^{\prime}(S^{\ast})D_{\theta}S^{\ast}[h] =0.
\end{equation*}
By symmetry of the second Frechet derivative, 
\begin{equation*}
D_{\theta\theta}^{2}V[h,k] = D_{\theta\theta}^{2}V[k,h].
\end{equation*}
Subtracting the second equality from the first therefore yields 
\begin{eqnarray*}
D_{\theta}V_y[k]h(S^{\ast}) + V_y h^{\prime}(S^{\ast})D_{\theta}S^{\ast}[k]
&=& D_{\theta}V_y[h]k(S^{\ast}) \\
&&+ V_y k^{\prime}(S^{\ast})D_{\theta}S^{\ast}[h].
\end{eqnarray*}

We next eliminate the terms involving $D_{\theta}V_y$. The functional Roy
identity in direction $k$ is 
\begin{equation*}
D_{\theta}V[k]+V_y k(S^{\ast})=0.
\end{equation*}
Differentiating this expression with respect to income $y$ gives 
\begin{equation*}
D_{\theta}V_y[k] + V_{yy}k(S^{\ast}) + V_y k^{\prime}(S^{\ast}) \frac{%
\partial S^{\ast}}{\partial y} =0,
\end{equation*}
and hence 
\begin{equation*}
D_{\theta}V_y[k] = - V_{yy}k(S^{\ast}) - V_y k^{\prime}(S^{\ast}) \frac{%
\partial S^{\ast}}{\partial y}.
\end{equation*}
Similarly, 
\begin{equation*}
D_{\theta}V_y[h] = - V_{yy}h(S^{\ast}) - V_y h^{\prime}(S^{\ast}) \frac{%
\partial S^{\ast}}{\partial y}.
\end{equation*}

Substituting these two expressions into the preceding equality, the terms
involving $V_{yy}h(S^{\ast})k(S^{\ast})$ cancel. Since $V_y>0$, division by $%
V_y$ yields 
\begin{eqnarray*}
h^{\prime}(S^{\ast}) \left\{ D_{\theta}S^{\ast}[k] + \frac{\partial S^{\ast}%
}{\partial y}k(S^{\ast}) \right\} &=& k^{\prime}(S^{\ast}) \left\{
D_{\theta}S^{\ast}[h] + \frac{\partial S^{\ast}}{\partial y}h(S^{\ast})
\right\}.
\end{eqnarray*}

Defining the compensated directional response in direction $k$ analogously
by 
\begin{equation*}
{\mathcal{S}}_k(y,\theta,\eta) = D_{\theta}S^{\ast}(y,\theta,\eta)[k] + 
\frac{\partial S^{\ast}(y,\theta,\eta)}{\partial y} k\left(S^{\ast}(y,%
\theta,\eta)\right),
\end{equation*}
the preceding equality can be written more compactly as 
\begin{equation}
h^{\prime}(S^{\ast}){\mathcal{S}}_k = k^{\prime}(S^{\ast}){\mathcal{S}}_h.
\end{equation}
Thus, the functional Slutsky symmetry condition follows from the functional
Roy identity together with symmetry of the second Frechet derivative of
indirect utility.

To see the connection with the finite-dimensional formulation, suppose 
\begin{equation*}
\theta (s)=P(s;\beta )
\end{equation*}%
and choose 
\begin{equation*}
h(s)=\frac{\partial P(s;\beta )}{\partial \beta _{j}},\qquad k(s)=\frac{%
\partial P(s;\beta )}{\partial \beta _{k}}.
\end{equation*}%
Then 
\begin{equation*}
h^{\prime }(S^{\ast })=P_{s\beta _{j}}(S^{\ast };\beta ),\qquad k^{\prime
}(S^{\ast })=P_{s\beta _{k}}(S^{\ast };\beta ),
\end{equation*}%
and 
\begin{equation*}
D_{\theta }S^{\ast }[h]=\frac{\partial S^{\ast }}{\partial \beta _{j}}%
,\qquad D_{\theta }S^{\ast }[k]=\frac{\partial S^{\ast }}{\partial \beta _{k}%
}.
\end{equation*}%
Hence the functional symmetry condition reduces to 
\begin{equation*}
P_{s\beta _{j}}\left\{ \frac{\partial S^{\ast }}{\partial \beta _{k}}+\frac{%
\partial S^{\ast }}{\partial y}P_{\beta _{k}}\right\} =P_{s\beta
_{k}}\left\{ \frac{\partial S^{\ast }}{\partial \beta _{j}}+\frac{\partial
S^{\ast }}{\partial y}P_{\beta _{j}}\right\} ,
\end{equation*}%
which is the finite-dimensional Slutsky symmetry condition.

\end{document}